\pdfoutput=1
\documentclass{lmcs}

\usepackage{lastpage}
\lmcsdoi{17}{3}{12}
\lmcsheading{}{\pageref{LastPage}}{}{}%
{Oct.~09,~2020}{Jul.~28,~2021}{}

\usepackage[utf8]{inputenc}
\usepackage{amsthm,amsmath,amssymb}
\usepackage{xargs,calc}
\usepackage{stmaryrd}
\usepackage{xcolor}
\usepackage{graphicx}
\usepackage{subfig}
\usepackage{boites}
\usepackage{xspace}
\usepackage{etoolbox}
\newlist{enuminline}{enumerate*}{2}
\setlist[enuminline]{label={\hspace{.5ex plus .3ex minus .2ex}\rm\arabic{*})},itemjoin*={{ }and{ }}}
\newlist{subthm}{enumerate}{3}
\setlist[subthm]{label=\enumstyle{\alph{*}},ref=\thethm\hspace*{1pt}\alph{*},%
  font=\normalfont,labelindent=*,leftmargin=*,start=1%
  ,widest=a}

\newcommand{\enumstyle}[1]{{\rm(#1)}}
\usepackage[bottom]{footmisc}

\graphicspath{{vcsg/}}

\usepackage{hyperref}

\newcommand{\AutomatonScale}{0.7}

\newcommand{\nplusun}{\hspace*{1.2pt plus 2pt}{+}\hspace*{0.6pt plus 2pt}1}
\newcommand{\nmoinsun}{\hspace*{1.2pt plus 2pt}{-}\hspace*{0.6pt plus 2pt}1}
\newcommand{\vmiminus}{\scalebox{0.75}[1.0]{\scriptsize$-$}}
\newcommand{\vmiplus}{\scalebox{0.75}[0.75]{\scriptsize$+$}}
\newcommand{\iplusun}{\hspace*{0.4pt}{\vmiplus}1}
\newcommand{\imoinsun}{\hspace*{0.4pt}{\vmiminus}1}

\newcommand{\po}{\mathchoice{+1}{\nplusun}{\iplusun}{+1}}
\newcommand{\mo}{\mathchoice{-1}{\nmoinsun}{\imoinsun}{-1}}

\renewcommand{\leq}{\leqslant}
\renewcommand{\geq}{\geqslant}
\renewcommand{\phi}{\varphi}
\renewcommand{\epsilon}{\varepsilon}
\renewcommand{\mod}{\text{~mod~}}

\newcommand{\strong}[1]{\textbf{#1}}
\newcommandx{\bfloor}[1]{\left\lfloor #1 \right\rfloor}
\newcommandx{\bceil}[1]{\left\lceil #1 \right\rceil}

\newcommand{\thmBlockFont}[1]{#1}
\theoremstyle{plain}
\newtheorem{claim}{\thmBlockFont{Claim}}[thm]
\renewcommand{\theclaim}{\thethm.\arabic{claim}}

\makeatletter 
 \newdimen\bk@hauteurcourrante
  \newdimen\bk@hauteursuivante
  \newdimen\bk@tempdim
\newenvironment{leftbar}{%
  \def\bk@espace{ }%
  \def\pt@to@bp##1{##1=.99627393548##1}
  \def\bkvz@before@breakbox{\ifhmode\par\fi\bk@hauteurcourrante=1200bp}%
  \def\bkvz@set@linewidth{\advance\linewidth-0.5\parindent}%
  \def\bkvz@left{\hskip 0.5pt\vrule\@width 0.5pt\hskip0.5\parindent\hskip -1pt}%
  \let\bkvz@right\relax
  \let\bkvz@top\relax
  \let\bkvz@bottom\relax
  \breakbox}{\endbreakbox}
\makeatother
\newenvironment{proofwithbar}[1][Proof of the claim]{\begin{leftbar}\noindent{\itshape#1}.~}{\end{leftbar}}
\newtheorem{falsetheoremX}{\thmBlockFont{Theorem}}
\newenvironment{falsetheorem}[1]{\renewcommand{\thefalsetheoremX}{\protect#1}\begin{falsetheoremX}}{\end{falsetheoremX}}

\newcommand{\lcorollary}[1]{\label{c.#1}}
\newcommand{\ldefinition}[1]{\label{d.#1}}

\newcommand{\llemma}[1]{\label{l.#1}}

\newcommand{\lproposition}[1]{\label{p.#1}}
\newcommand{\lproperty}[1]{\label{pp.#1}}

\newcommand{\lsection}[1]{\label{s.#1}}

\newcommand{\lfigure}[1]{\label{f.#1}}
\newcommand{\ltheorem}[1]{\label{t.#1}}
\newcommand{\lequation}[1]{\label{eq.#1}}
\newcommand{\lclaim}[1]{\label{cl.#1}}
\newcommand{\preprocgenref}[2]{}
\newcommand{\generalref}[2]{%
  \preprocgenref{#1}{#2}%
  \ifthenelse{\equal{#1}{eq}}%
  {(\ref{#1.#2})}%
  {\ref{#1.#2}}%
}
\newcommand{\generalpageref}[2]{\pageref{#1.#2}}

\makeatletter
\newcommand*{\ralgorithm}{\@ifstar{\generalref{a}}{Algorithm~\ralgorithm*}}
\newcommand*{\palgorithm}{\@ifstar{\generalpageref{a}}{page~\palgorithm*}}

\newcommand*{\rcorollary}{\@ifstar{\generalref{c}}{Corollary~\rcorollary*}}
\newcommand*{\pcorollary}{\@ifstar{\generalpageref{c}}{page~\pcorollary*}}

\newcommand*{\rdefinition}{\@ifstar{\generalref{d}}{Definition~\rdefinition*}}
\newcommand*{\pdefinition}{\@ifstar{\generalpageref{d}}{page~\pdefinition*}}

\newcommand*{\rexample}{\@ifstar{\generalref{e}}{Example~\rexample*}}
\newcommand*{\pexample}{\@ifstar{\generalpageref{e}}{page~\pexample*}}

\newcommand*{\rlemma}{\@ifstar{\generalref{l}}{Lemma~\rlemma*}}
\newcommand*{\plemma}{\@ifstar{\generalpageref{l}}{page~\plemma*}}

\newcommand*{\rproblem}{\@ifstar{\generalref{pb}}{Problem~\rproblem*}}
\newcommand*{\pproblem}{\@ifstar{\generalpageref{pb}}{page~\pproblem*}}

\newcommand*{\rproposition}{\@ifstar{\generalref{p}}{Proposition~\rproposition*}}
\newcommand*{\pproposition}{\@ifstar{\generalpageref{p}}{page~\pproposition*}}

\newcommand*{\rproperty}{\@ifstar{\generalref{pp}}{Property~\rproperty*}}
\newcommand*{\pproperty}{\@ifstar{\generalpageref{pp}}{page~\pproperty*}}

\newcommand*{\rprocedure}{\@ifstar{\generalref{pc}}{Procedure~\rprocedure*}}
\newcommand*{\pprocedure}{\@ifstar{\generalpageref{pc}}{page~\pprocedure*}}

\newcommand*{\rremark}{\@ifstar{\generalref{r}}{Remark~\rremark*}}
\newcommand*{\premark}{\@ifstar{\generalpageref{r}}{page~\premark*}}

\newcommand*{\rnotation}{\@ifstar{\generalref{n}}{Notation~\rnotation*}}
\newcommand*{\pnotation}{\@ifstar{\generalpageref{n}}{page~\pnotation*}}

\newcommand*{\rsection}{\@ifstar{\generalref{s}}{Section~\rsection*}}
\newcommand*{\psection}{\@ifstar{\generalpageref{s}}{page~\psection*}}

\newcommand*{\rtable@}{\@ifstar{\generalref{t}}{Table~\rtable*}}
\newcommand*{\rtable}{\protect\rtable@}
\newcommand*{\ptable}{\@ifstar{\generalpageref{t}}{page~\ptable*}}

\newcommand*{\rfigure}{\@ifstar{\generalref{f}}{Figure~\rfigure*}}
\newcommand*{\pfigure}{\@ifstar{\generalpageref{f}}{page~\pfigure*}}

\newcommand*{\requation}{\@ifstar{\generalref{eq}}{Equation~\requation*}}
\newcommand*{\pequation}{\@ifstar{\generalpageref{eq}}{page~\pequation*}}

\newcommand*{\rtheorem}{\@ifstar{\generalref{t}}{Theorem~\rtheorem*}}
\newcommand*{\ptheorem}{\@ifstar{\generalpageref{t}}{page~\ptheorem*}}

\newcommand*{\rclaim}{\@ifstar{\generalref{cl}}{Claim~\rclaim*}}
\newcommand*{\pclaim}{\@ifstar{\generalpageref{cl}}{page~\pclaim*}}
\makeatother

\newcommand{\halfspace}{\hspace{0.5\fontdimen2\font plus 0.5\fontdimen3\font
minus 0.5\fontdimen4\font}}
\newcommand{\thirdspace}{\hspace{0.33\fontdimen2\font plus 0.33\fontdimen3\font
minus 0.33\fontdimen4\font}}

\makeatletter
\newlength{\vm@xmd@d}
\newlength{\vm@xmd@n}
\newlength{\vm@xmd@s}
\newlength{\vm@xmd@ss}
\newcommand{\xmd}{%
  \ifmmode%
    \mathchoice%
    {\hspace*{\vm@xmd@d}}
    {\hspace*{\vm@xmd@n}}
    {\hspace*{\vm@xmd@s}}
    {\hspace*{\vm@xmd@ss}}
  \else%
    \hspace*{\vm@xmd@n}%
  \fi%
}
\makeatother

\newcommand{\card}[1]{{\tt Card}(#1)}
\newcommandx{\wlen}[1]{|#1|}
\newcommand{\Z}{\mathbb{Z}}
\newcommand{\N}{\mathbb{N}}
\newcommand{\Q}{\mathbb{Q}}
\newcommand{\R}{\mathbb{R}}
\newcommand{\lang}{\textsc{Lang}}
\newcommand{\pref}[1]{\text{Pref}\,(#1)}
\newcommand{\set}[1]{%
  \left\{\mathchoice%
  {\halfspace #1 \halfspace}%
  {\thirdspace #1 \thirdspace}%
  {#1}%
  {#1}\right\}%
}
\newcommand{\setq}[2]{\left\{\halfspace#1~\middle|~#2\halfspace\right\}}

\newcommand{\arobase}{\raisebox{-0.2ex}{\fontfamily{ptm}\selectfont @}}
\newcommand{\mel}[2]{{\tt\href{mailto:#1@#2}{#1\arobase{}#2}}}

\newcommand{\mathsc}[1]{{\normalfont\textsc{#1}}}

\let\val\relax
\DeclareMathOperator{\rep}{\mathsc{Rep}}
\DeclareMathOperator{\val}{\mathsc{Val}}
\DeclareMathOperator{\valfk}{\mathsc{Val}^{\text{\cite{FrouKlou12}}}}
\DeclareMathOperator{\repfk}{\mathsc{Rep}^{\text{\cite{FrouKlou12}}}}

\newcommand{\Npq}{N_{\frac{p}{q}}}
\newcommand{\Npqfk}{N_{\frac{p}{q}}^{\text{\cite{FrouKlou12}}}}
\newcommand{\pq}{\frac{p}{q}}
\newcommand{\Vpq}{V_{\frac{p}{q}}}
\newcommand{\Vpqp}{W_{\frac{p}{q}}}

\newcommand{\digit}{\textsf{digit}}

\newcommand{\bpq}{\beta}

\newcommand{\fo}[2][]{{\tt FO}_{#1}\left[#2\right]}
\newcommand{\vect}[1]{\overline{#1}}
\newcommand{\logic}[1][]{\fo[#1]{\Npq,+,\Vpq}}
\newcommand{\Npqd}[1][d]{\big(\Npq\big){}^{#1}}

\newcommand{\defeq}{\mathrel{:=}}

\newcommand{\pqdef}{$\pq$-definable\xspace}
\newcommand{\pqrec}{$\pq$-recognisable\xspace}
\newcommand{\pqrecy}{$\pq$-recognisability\xspace}
\newcommand{\pqrecs}{$\pq$-recognisable\xspace}

\newcommand{\pqreps}{$\pq$-representations\xspace}

\newcommand{\suff}[1]{\mathsc{Suff}_{#1}}
\renewcommand{\pref}[1]{\mathsc{Pref}_{#1}}

\newcommand{\Ac}{\mathcal{A}}
\newcommand{\Bc}{\mathcal{B}}
\newcommand{\Cc}{\mathcal{C}}

\newcommand{\Ap}{A_p}
\newcommand{\Aps}{\left(A_p\right)^*}
\newcommand{\Apd}[1][d]{\Ap^{\hspace*{2pt}#1}}
\newcommand{\Apds}[1][d]{\left(\Apd[#1]\right)^*}

\newcommand{\tuple}[1]{\mathbf{u}}

\newcommand{\Lpq}{L_\pq}
\newcommand{\Lpqfk}{L_{\pq}^{\text{\cite{FrouKlou12}}}}

\newcommand{\ie}{\textit{i.e.\@}\xspace}
\newcommand{\eg}{\textit{e.g.\@}\xspace}
\newcommand{\cf}{\textit{cf.}~}

\newcommand{\Hsquare}{%
  \text{\fboxsep=-.2pt\fbox{\rule{0pt}{1.25ex}\rule{1.5ex}{0pt}}}
}

\newcommand{\relvee}{\mathrel{\vee}}
\newcommand{\relwedge}{\mathrel{\wedge}}

\newcommand{\lord}[1]{%
  \ifthenelse{\equal{#1}{<}}{\mathrel{\vartriangleleft}}{%
  \ifthenelse{\equal{#1}{>}}{\mathrel{\vartriangleright}}{%
  \ifthenelse{\equal{#1}{\leq}}{\mathrel{\trianglelefteq}}{%
  \ifthenelse{\equal{#1}{\geq}}{\mathrel{\trianglerighteq}}{%
  \ifthenelse{\equal{#1}{=}}{\mathrel{\Hsquare}}{%
    \lord\leq#1%
  }}}}}%
}

\newcommand{\fcdots}{\cdot{\cdot}\cdot}

\newcommand{\pqp}{\big(\pq\big)}

\newcommand{\logimplies}{\rightarrow}

\makeatletter
\newdimen\wof@@length
\newcommand{\WidthOf}[1]{%
  \setbox0=\hbox{#1}
  \wof@@length=\wd0
  \the\wof@@length
}
\makeatother

\DeclareMathSymbol{\in}{\mathbin}{symbols}{"32}
\newcommand{\noneqspacing}{
  \thickmuskip=5mu plus 3mu minus 2mu
  \medmuskip=3mu plus 2mu minus 2mu
}
\noneqspacing
\newcommand{\eqspacing}{
  \thickmuskip=12mu plus 3mu minus 3mu
  \medmuskip=5mu plus 3mu minus 2mu
}%
\AtBeginEnvironment{gather*}{\eqspacing}%
\AtBeginEnvironment{gather}{\eqspacing}%
\AtBeginEnvironment{equation}{\eqspacing}%
\AtBeginEnvironment{equation*}{\eqspacing}%
\AtBeginEnvironment{align}{\eqspacing}%
\AtBeginEnvironment{align*}{\eqspacing}%
\AtBeginEnvironment{multline}{\eqspacing}%
\AtBeginEnvironment{multline*}{\eqspacing}%

\title{On p/q-recognisable sets}
\author[V.~Marsault]{Victor Marsault}
\address{LIGM, Université Gustave Eiffel, CNRS, F-77454 Marne-la-Vallée, France}
\address{Laboratory for Foundations of Computer Science, School of Informatics, University of Edinburgh, United Kingdom}
\address{}

\email{\mel{victor.marsault}{univ-eiffel.fr}}
\urladdr{\url{http://victor.marsault.xyz}}
\begin{document}

\maketitle

\begin{abstract}\renewcommand{\Npq}{N_{\raisebox{.2ex}[\height][\depth-.1ex]{\hspace*{-.1ex}$\scriptstyle\pq$}}}
  Let~$\frac{p}{q}$ be a rational number.
  Numeration in base~$\pq$ is defined by a function that evaluates each finite
  word over~$A_p=\{0,\,1,\,\ldots,\allowbreak{p-1}\}$ to a rational number.
  We let~$\Npq$ denote the image of this evaluation function.
  In particular,~$\Npq$ contains all nonnegative integers and the literature on
  base~$\pq$ usually focuses on the set of words that are evaluated to
  nonnegative integers; it is a rather chaotic language which is not
  context-free.
  On the contrary, we study here the subsets of~$(\smash{\Npq})^d$ that are
  \pqrecs, \ie realised by finite automata over~$\left(\Ap\right)^d$.
  First, we give a characterisation of these sets as those definable in a
  first-order logic, similar to the one given by the B{\"u}chi-Bruy{\`e}re
  Theorem for integer base numeration systems.
  Second, we show that the natural order relation and the modulo-$q$ operator
  are not \pqrec.  
\end{abstract}

\maketitle

\section{Introduction}

Let $p$ and~$q$ be two coprime integers such that~$p>q>1$, hence~$\frac{p}{q}$
is an irreducible fraction greater than $1$.
The base-$\frac{p}{q}$ (numeration system) was introduced in~\cite{AkiyEtAl08}.
The study of such \emph{rational base numeration systems} have enabled progress
to be made in solving deep problems from number theory (Josephus
and Mahler problems \cite{Akiy08}).
Like other numerations systems, base~$\frac{p}{q}$ gives a way to
\emph{represent} numbers by words, and to \emph{evaluate} words to numbers.
Here, we consider only finite words over the canonical
alphabet~$\Ap=\set{0,1,\ldots,p-1}$; it is the smallest digit-set $X$
such that all nonnegative integers have representations using only digits
from~$X$.
One paradox of base~$\frac{p}{q}$ is that simple number-sets are
represented by complicated languages and that simple languages are evaluated to
complicated number sets.
For instance,~$\N$ is represented by a rather chaotic language,~$\Lpq$,
that does not fit well in the usual hierarchy of formal languages.
On the other hand, the evaluations of all words over~$\Ap$ form a set of
numbers,~$\Npq$, which is hard to describe arithmetically.
Literature on base~$\pq$ mostly focuses on~$\Lpq$.
It is not a context-free language \cite{AkiyEtAl08}
and even defeats any kind of iteration lemma~\cite{MarsSaka13b}.
However, there is a very simple periodic procedure to generate $\Lpq$ that is
similar to a breadth-first search \cite{MarsSaka17b}.
Some effort have also been made to study~$\Lpq$ from the perspective of
combinatorics on words
(frequency of patterns, sum-of-digit function, \textit{etc.}
\cite{MorgEtAl14,EdgaEtAlXX_18,Dubi09}).
On the contrary, there has been little work on~$\Npq$.
Here, we start exploring this area by studying the subsets of~$\Npqd$ that are
realised by finite automata reading synchronously on~$d$ tapes.
We call such sets \emph{\pqrec}.
In the original article introducing base~$\pq$ \cite{AkiyEtAl08}, it is shown
that addition is \pqrec (in fact, normalisation of representations from any
finite alphabet).
This is a strong property which, in the case of integer base~$b$ leads to the
following characterisation of $b$-recognisable sets.
%

\begin{thm}[Büchi--Bruyère Theorem \cite{BruyEtAl94}]\ltheorem{buch-bruy}
  A subset of~$\N^d$ is~$b$-recognisable if and only if it is definable in the
  first-order logic~$\fo{\N,+,V_b}$, where~$V_b$ is the function that maps~$n$
  to the greatest power of~$b$ that divides~$n$.  
\end{thm}

Similar statements are also known for other numeration systems
(\eg \emph{Pisot U-systems}, \emph{Pisot~$\beta$-numerations}) and in other
settings (\eg characterising the sets of infinite words realised by Büchi
automata) \cite{Char18-ib}.
In the present article, we show such a logic characterisation of \pqrec sets,
stated below.
\begin{thm}\ltheorem{pq-defi<->pq-reco}
  A subset of~$\Npqd$ is \pqrec if and only if it is definable in the
  first-order logic 
    $\logic$,
  where
    $\Vpq$ is the function~$\Npq\rightarrow\Npq$ that maps~$x$ to the greatest
    power of~$\pq$ that one may divide~$x$ by and obtain a quotient in~$\Npq$.
\end{thm}
It could seem strange that in our case, the universe is~$\Npq$
instead of~$\N$.
However, since~$\Lpq$ is not a regular language,
the classical generalisation of Theorem 1 to rational base would clearly
not hold.
Note also that something is hidden in this statement.
In most numerations systems, \emph{natural order} may be expressed using
\emph{addition} ($y\leq z$ may be expressed by~$\exists x,~x+y=z$) and the
proofs of the logic characterisations (\eg the one of \rtheorem{buch-bruy})
rely on this fact.
On the contrary, this is not true in base~$\pq$ (see below) and we introduce
another pre-order relation (denoted by $\lord$) in our proof.

\medskip

Very little is known about the expressive power of automata with respect to
base~$\frac{p}{q}$.
In section \ref{s.nota}, we begin to investigate this area.
First, we show that the natural order relation over $\Npq$ is not \pqrec.
This is quite surprising since
\begin{enuminline}
  \item the contrary is true in almost all numeration
systems, 
  \item the natural order relation over $\N$ is recognised by a trivial finite
automaton, provided that we are ensured that input words belong to~$\Lpq$.
\end{enuminline}
Second, we consider the modulo-$n$ operator, where~$n$ is an integer constant.
This class of operators is related to the periodic subsets of~$\N$ which,
since the work of Cobham, have a particular place within the study of
numeration systems.
In fact, most numeration systems~$S$ are such that 
\begin{enuminline}
  \item $\N$
is~``$S$-recognisable'',
  \item periodic subsets of~$\N$ are $S$-recognisable.
\end{enuminline}
(These properties hold if~$S$ belongs to the very large class of
\emph{regular abstract numeration systems}~\cite{LecoRigo10-ib}.)
In base~$\frac{p}{q}$, neither Item~1 nor~2 hold;
it is not surprising since the relevant set of numbers is~$\Npq$ rather
than~$\N$.
By definition, the set~$\Npq$ is \pqrec, and the last part of this work is about
adapting Item~2.
If~$n$ is coprime with~$q$, then the modulo-$n$ operator is easy
to generalise as a function ${\Npq\rightarrow\Z/n\Z}$ and we show that it is 
\pqrec (\rcorollary{pqrec.modu}).
On the other hand, when~$n$ is not coprime with~$q$,
no generalisation of the modulo-$n$ operator is obvious.
Nevertheless, we show that there is no \pqrec set that separates~$q\N$ from
its complement in~$\N$
(\rproposition{npqrec.modu}).
It follows that no generalisation of the modulo-$n$ operator
would be \pqrec if~$n$ is a multiple of~$q$ (\rcorollary{npqrec.modu}).
%

\section{Preliminaries}

\subsection{Words, automata}
An \emph{alphabet}~$A$ is a finite set of symbols, called indifferently
\emph{letters} or \emph{digits}.
A \emph{word} over~$A$ is a finite sequence $u=a_{k\mo}\cdots a_1a_0$ of
letters from~$\Ac$.
We let~$\wlen{u}$ denote the \emph{length} of~$u$, that is $\wlen{u}=\wlen{a_{k\mo}\cdots a_1a_0}=k$.
For each~$i$,~$0\leq i\leq\wlen{u}$, we denote the prefix of~$u$ of length~$i$
by~$\pref{i}(u)$ that is~$\pref{i}(u)=a_{k\mo}\ldots a_{k-i}$;
similarly~$\suff{i}(u)=a_{i\mo}\ldots a_{0}$ denotes the suffix of~$u$ of
length~$i$.
We denote the set of all words over~$A$ by~$A^*$ and
subsets of~$A^*$ are called \emph{languages over~$A$}.
The set~$A^*$ is endowed with the
\emph{concatenation}, usually denoted implicitly as in~$u\xmd v$,
or explicitly by a middle dot when it helps readability, as in~$u\cdot v$.
A \emph{(deterministic) automaton}~$\Ac$ is defined by a 5-tuple~$\Ac=\langle\,A,\,
Q_\Ac,\, i_\Ac, \delta_\Ac,F_\Ac \rangle$, where~$A$ is an alphabet,~$Q_\Ac$ is
a finite set of \emph{states}, $i_\Ac\in Q_\Ac$ is the \emph{initial
state},~$\delta_\Ac:Q_\Ac\times A \rightarrow Q_\Ac$ is the (partial)
\emph{transition function} and~$F_\Ac\subseteq Q_\Ac$ is a set of \emph{final
states}.
As usual, we extend~$\delta_\Ac$ as a function~$Q_\Ac\times A^*\rightarrow
Q_\Ac$ by~$\delta_\Ac(q,\epsilon)=q$ and~$\delta_\Ac(q,a\xmd
u)=\delta_\Ac(\delta_\Ac(q,a),u)$.
We call \emph{the run of~$u$}, if it exists, the sequence of states
reached during the execution of~$\Ac$ on~$u$, namely, the finite
sequence~$\left(\delta_{\Ac}(i_\Ac,\pref{i}(u))\right)_{0\leq i \leq \wlen{u}}$;
in particular, the phrase \emph{the run of~$u$ reaches state~$s$}
means that $\delta_\Ac$ is defined on $(i_\Ac,u)$ and
that~$\delta_\Ac(i_\Ac,u)=s$.
A word~$u$ is said to be \emph{accepted} by~$\Ac$ if its run exists and
reaches a final state; the \emph{accepted language} of~$\Ac$, denoted
by~$\lang(\Ac)$
is the set of the words accepted by~$\Ac$.
An automaton is \emph{complete} if its transition function is total.
It is sometimes more convenient to have an automaton read number
representations \emph{most} significant digit first (MSDF), and sometimes
\emph{least} significant digit first (LSDF).
Since the position of a digit is meaningful for evaluation, we try to keep
consistent the indexing of words throughout the article, hence we want to avoid
mirroring words and languages.
Thus, we say that an automaton is \emph{left-to-right} if it reads its input in
a normal fashion, and \emph{right-to-left}  if it reads
its input from right to left.
Since we usually represent numbers MSDF (\cf Section~\ref{s.defi-rati-base}),
\emph{left-to-right} automata work on MSDF representations, while
\emph{right-to-left} automata work on LSDF representations.
In a right-to-left automaton, the function~$\delta_\Ac$ is actually generalised
as~$\delta_\Ac(q,u\xmd a)=\delta_\Ac(\delta_\Ac(q,a),u)$, and the run of a word~$u$
refers to the state sequence
$\left(\delta_{\Ac}(i_\Ac,\suff{i}(u))\right)_{0\leq i\leq \wlen{u}}$.

We will always consider automata in relation with a rational base numeration
system, hence automata will all be over the alphabet~$(\Ap)^d$ for some
\emph{number~$d$ of tapes}.
(Rational bases and the digit set~$\Ap$ are defined in \rsection{defi-rati-base}.)
A word~$\vect{u}$ in~$\Apds$ may be divided component-wise as~$\vect{u}=(u_0,u_1,\ldots,u_{d\mo})$
where~$u_0,u_1,\ldots,u_{d\mo}$ are words in~$\Aps$ that are of equal length.
The digit~$0$ always belongs to~$\Ap$ and when~$d$ is clear from context,
we let~$\vect{0}$ denote the letter~$(0,\ldots,0)$ which is part of~$\Apd$.
Moreover, we say that a~$d$-tape automaton is \emph{padded} if for every
word $\vect{u}$ in~$\Apds$,
$\vect{u}$ is accepted by~$\Ac$ if and only if~$\vect{0}\cdot \vect{u}$
is also accepted.
Intuitively, a padded automaton accepts words \emph{by value};
indeed, the~$i$-th component of $\vect{u}$ and of~$\vect{0}\cdot \vect{u}$
have the same value.
In the following, we will exclusively consider automata
that are padded.
\subsection{First-order logic}
We briefly recall the definition of first-order formulas.
Let~$X$ be a countable set of \emph{variables}.
Let~$\Omega$ be a set called the \emph{domain},
let~$\big(f_i\big)_{i\in I}$ be a family of functions,
let~$\big(R_j\big)_{j\in J}$ be a family of relations on~$\Omega$,
and~$\big(c_k\big)_{k\in K}$ be a family of constants in~$\Omega$.
We 
let
\begin{equation}
  \fo{\Omega,\,\big(f_i\big)_{i\in I},\,\big(R_j\big)_{j\in J},\,(c_k\big)_{k\in K}}
\end{equation}
denote the logic in which terms and formulas are
defined recursively as follows.
\begin{subthm}[label=\textbullet]
  \item Each constant $c_k$, with~$k\in K$, and each variable~$x\in X$ is a term.
  \item If~$f_i$ is a~$n$-ary function and~$t_0, \ldots, t_{n\mo}$ are terms,
        then $f_i(t_0, \ldots, t_{n\mo})$ is a term.
  \item If~$t$ and~$t'$ are two terms, then~$t=t'$ is a formula.
  \item If~$R_j$ is a~$n$-ary relation and~$t_0, \ldots, t_{n\mo}$ are terms,
    then $R_j(t_0, \ldots, t_{n\mo})$ is a formula.
  \item If~$\phi$ is a formula, then~$\neg\phi$ is a formula.
  \item If~$\phi$ and~$\psi$ are formulas, then~$\phi \relwedge\psi$,~$\phi\relvee \psi$ and~$\phi \logimplies \psi$ are formulas.
  \item If~$\phi$ is a formula and~$x$ is a variable, then~$\exists x~\phi$ and~$\forall x~\phi$ are formulas
\end{subthm}
A variable~$x$ is called \emph{free} in a formula~$\phi$ if it appears outside
the scope of any quantifier~$\exists x$ or~$\forall x$.
If a formula~$\phi$ has~$d$ free variables, we usually make them explicit
by writing~$\phi$ as~$\phi(x_0,x_1,\ldots,x_{d\mo})$.
We do not define in all generality the interpretation of relations, functions
and formulas since we will always use the straightforward interpretations.
Hence we assume that it is clear what it means for a closed formula
to be true and we say that a subset~$S$ of~$\Omega^d$ is \emph{defined by the formula~$\phi$} if the following holds.
\begin{equation}
  S=\setq{ \vect{z}\in \Omega^d}{\phi(\vect z)\text{ is true in }\Omega}
\end{equation}

\subsection{Rational base numeration systems}
\lsection{defi-rati-base}

\medskip

Let~$p$ and~$q$ be two coprime integers such that~$p>q>1$; they will
be fixed throughout the article.
Note that~$\frac{p}{q}$ is thus an irreducible fraction greater than 1.
We define below base-$\pq$ (numeration system) ;
for more details see~\cite{FrouSaka10-ib,Mars16}.
Let us stress that it is \strong{not} the special case of a
\emph{$\beta$-numeration} where~$\beta$ is a rational number.

\subsubsection{Evaluation}%
The \emph{canonical alphabet} associated with base~$\pq$, denoted by~$\Ap$, is
the set~${\Ap=\set{0,1,\ldots,p\mo}}$.
The \emph{evaluation function}  maps every word~$u=a_{k-1}\cdots a_1a_0$
in~$\Aps$ to a rational number as follows.
\begin{equation}\lequation{eval}
  \val(u)=\val(a_{k-1}\cdots a_1 a_0) =
  \sum^{k-1}_{i=0} \frac{a_i}{q} \left( \pq \right)^i
\end{equation}
The number $\val(u)$ is called \emph{the value of $u$}.
Note that the value of a word may be computed recursively: for every~$u,w\in\Aps$,
the following holds.
\begin{subequations}\lequation{eval-recu}
\begin{align}
  \val(u\xmd w)={}& \val(u\xmd 0^{\wlen{w}})+ \val(w) \\
  ={}& \val(u)\left(\pq\right)^{\wlen{w}} + \val(w)
\end{align}
\end{subequations}

\begin{nota}
  We let~$\Npq$ denote the image of the evaluation
  function:~$\Npq=\val\left(\Aps\right)$.
\end{nota}

\begin{figure}
  \centering
  \includegraphics[width=\linewidth]{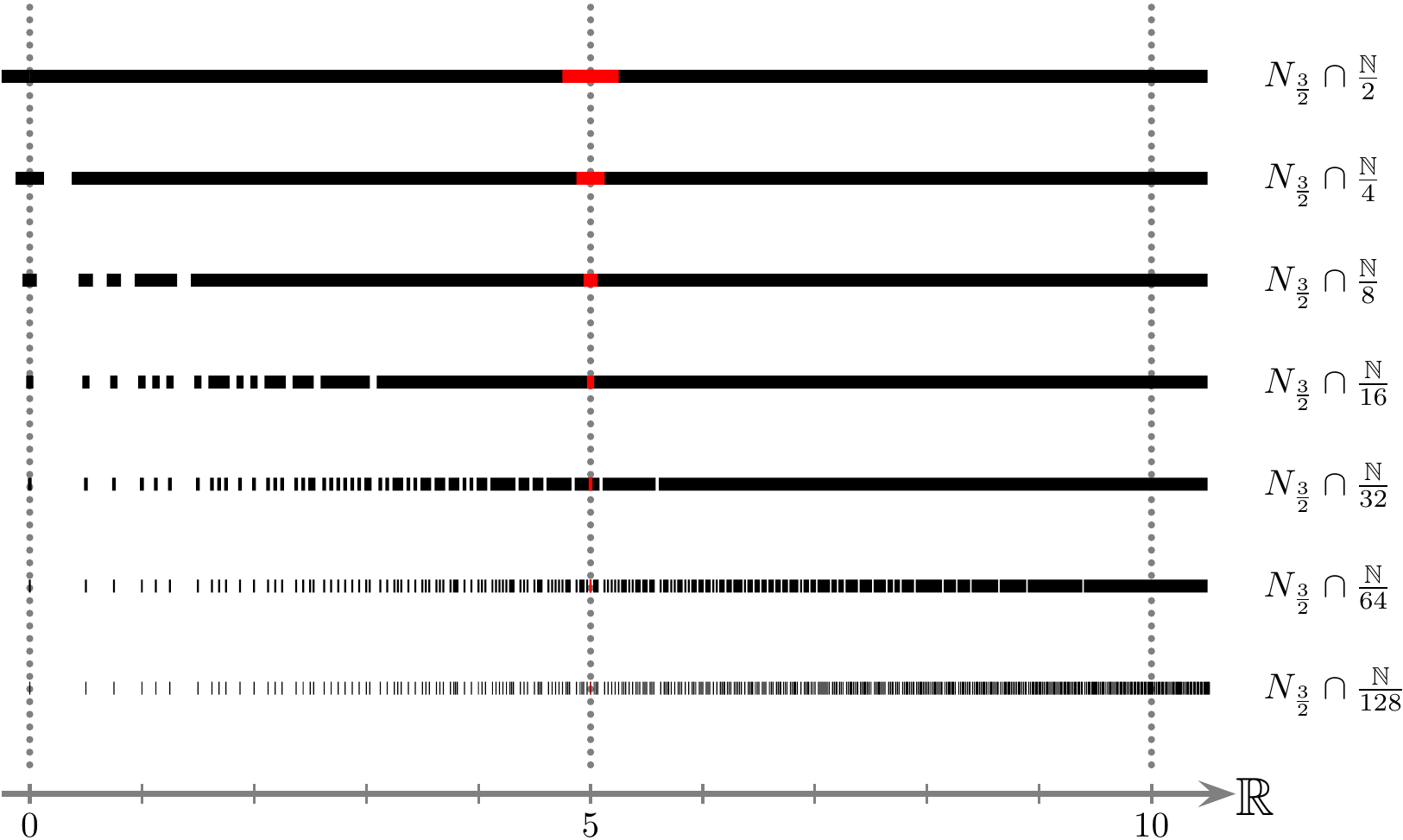}
  \caption{$N_\frac{3}{2}$, built by successive refinements}
  \lfigure{v32}
\end{figure}

The set~$\Npq$ is hard to describe in another way than its definition, and in
general little is known about it.
Of course, \requation{eval} implies that $\Npq$ contains only rational numbers
that have for denominator a power
of~$q$.
Moreover, one may show that for a given power~$q^k$, only
finitely many numbers in~$\frac{\N}{q^k}$ do not belong to~$\Npq$ (see
\cite{MarsSaka13b} or \rlemma{vpq-ulti-all}, later on).

\begin{exa}
  \rfigure{v32} gives an intuition of the content of~$\Npq$ for
  base~$\frac{3}{2}$:
  the~$k$-th line represent $\Npq\cap\frac{\N}{q^k}$ and each
  number~$\frac{n}{q^k}$ in~$\Npq$ is represented by a black segment of
  length~$\frac{1}{q^k}$ centred around abscissa~$\frac{n}{q^k}$;
  in each line, the highlighted segment always represents the number~$5$.
  The figure highlights the fact that as we refine the approximation of~$\Npq$,
  more and more holes appear in a seemingly chaotic fashion.
\end{exa}
%

%
\subsubsection{Representation}
Given a number~$x$, we call \emph{expansion of~$x$} any word~$u$ in~$\Aps$ such
that~$\val(u)=x$.
By definition, each number in~$\Npq$ has at least one expansion, and it is known
that it is unique up to leading~$0$'s:

\begin{thmC}[\cite{AkiyEtAl08}]\ltheorem{repr-uniq}
  If two words~$u$ and~$v$ in~$\Aps$ are such that~$\val(u)=\val(v)$
  and ${\wlen{u}\leq \wlen{v}}$,
  then~$0^i\xmd u= v$ for some nonnegative integer~$i$.
\end{thmC}

We call \emph{representation of~$x$} and let~$\rep(x)$ denote the unique
expansion that does not start by the digit~$0$.
Similarly, a $d$-tuple $\vect{x}=(x_0,\ldots,x_{d\mo})$ in~$\Npqd$
is represented by the unique word $\rep(\vect{x})=\vect{u}=(u_0,\ldots,u_{d\mo})$ in~$\smash{\Apds}$
that does not start by~$\vect{0}$ and
such that every~$u_i$ belongs to~$0^*\rep(x_i)$.
%

%
Every integer belongs to~$\Npq$ and its representation may be computed by the following right-to-left
algorithm.%
\footnote{An algorithm computing~$\rep(x)$, for~$x$ in~$\Npq$ and beyond, is given in \cite{FrouKlou12}.}
%
\begin{subequations}\lequation{MEA-alt}%
  \begin{align}
  \lequation{MEA-alt-1}
    \rep(0) ={}& \epsilon
    \\[1mm]
    \forall m\in\N,~m\mathbin{>}0\quad
    \rep(m) ={}&  \rep(n) \cdot  a
    \lequation{MEA-alt-2}
      \quad\quad\text{where }\left\{
      \begin{array}{l}
        n\in \N,~ a\in\Ap \\
        q\xmd m = p\xmd n +a
      \end{array}\right.
  \end{align}
\end{subequations}
%
%
\begin{nota}
  We let~$\Lpq$ denote the set of the representations of nonnegative integers, that is ${\Lpq=\rep(\N)}$
\end{nota}
As briefly recalled in the introduction,~$\Lpq$ has been well studied; in the
present work we will use only the following two properties:~$\Lpq$ is prefix-closed
(follows from~\requation*{MEA-alt}) and it is not a regular language
(\cf \cite{FrouSaka10-ib,Mars16}).
Note that~$0^*\Lpq$ is also a prefix-closed language that is not regular.
\newcommand{\vmrule}{\rule[-1.4ex]{0ex}{3.5ex}}
\subsubsection{\protect\pqrec sets}
Let us now define for base~$\pq$ a notion analogous to the one
of~$b$-recognisable sets defined in the context of an integer base~$b$
\cite{BruyEtAl94}.

\begin{defi}
  Let~$d$ be a positive integer and~$S$ a subset of~$\Npqd$.
  \begin{subthm}
    \item A~$d$-tape automaton~$\Ac$ is said to \emph{realise}~$S$
    if~$\lang(\Ac)=\left(\vect{0}\right)^{\hspace*{-2pt}*} \rep(S)$.
    \item $S$ is said \emph{\pqrec} if it is realised by some automaton.
  \end{subthm}
\end{defi}

Note that item \enumstyle{a} implies in particular that~$\Ac$ is padded and
that, since the expansions of a number are equal up to leading~$0$'s,~$\Ac$
accepts by value:~$\lang(\Ac)=\val^{-1}(S)$.

\begin{exa}
  $\Npq$ is \pqrec since~$\val^{\mo}(\Npq)=\Aps$,
  but~$\N$ is not since~$\val^{\mo}(\N)=0^*\xmd\Lpq$ is not a regular language.
\end{exa}

\begin{thmC}[\cite{AkiyEtAl08}]\ltheorem{addi}
 Addition is \pqrec.
\end{thmC}

\begin{figure}
  \centering
  \includegraphics[scale=\AutomatonScale]{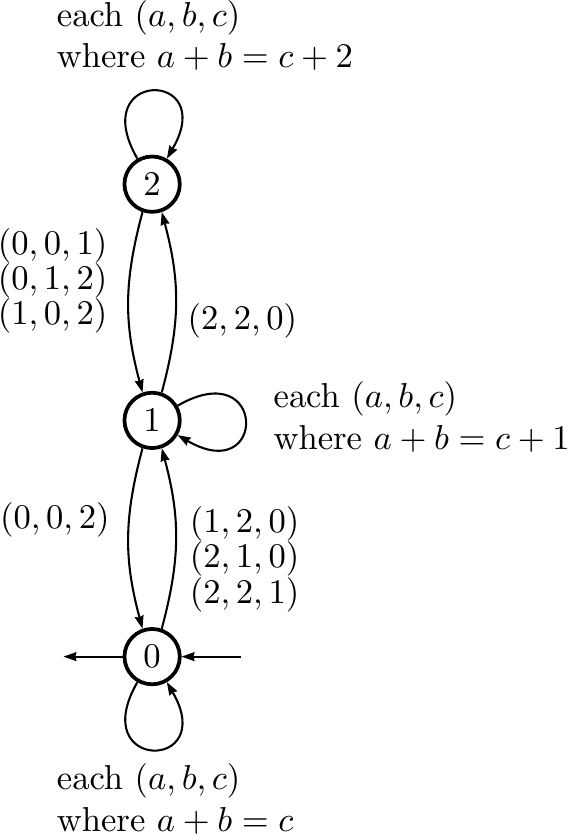}
  \caption{$\Cc_\frac{3}{2}$, a 3-tape right-to-left automaton that realises addition in base~$\frac{3}{2}$}
  \lfigure{add_3_2}
\end{figure}

Since it is central to our logic characterisation, we give without proof the
definition of~$\Cc_\pq$, the 3-tape right-to-left automaton that realises
addition in base~$\pq$.
\begin{subequations}\lequation{addi}
\begin{equation}
  \Cc_\pq = \left\langle~\Apd[3],~Q_\Cc,~i_\Cc,~\delta_\Cc,~F_\Cc~\right\rangle~,
\end{equation}
where the state set is~$Q_\Cc = \set{0,\ldots, m }$ with~$m \mathbin{=} 1+\bfloor{\frac{p-2}{p-q}}$;
the initial state is $i_\Cc = 0$; the final-state set is $F_\Cc=\set{0}$; and
the transition function is defined as follows.
  \begin{multline}
    \forall s\mathop{\in} Q_\Cc,~\forall(a,b,c)\mathop{\in}\Apd[3]\quad\quad
    \delta_\Cc(s,(a,b,c)) \mathbin{=} s'\quad\quad\text{if}\quad q\xmd s\mathop{+}a\mathop+b \mathbin{=} p\xmd s' \mathop+ c
  \end{multline}
\end{subequations}
For instance, \rfigure{add_3_2} shows~$\Cc_\pq$ for base~$\frac{3}{2}$.
Note that~$\delta_\Cc$ is not a total function.
In fact, if~$s$,~$a$ and~$b$ are fixed, there is exactly one~$c$ in~$\Ap$ such
that $\delta_\Cc(s,(a,b,c))$ is defined.
Hence, for each length~$i$, there is exactly word $u_i$ of length~$i$
such that $\delta_{\mathcal C}\big(s,(0^i,0^i,u_i\big)$ is defined.
Moreover, if~$i\geq m$, it may be shown 
that~$\delta_{\mathcal C}\big(s,(0^i,0^i,u_i\big)=0$.
In particular, the following holds.

\begin{pty}
  $\Npq$ is stable by addition :\quad
    $\displaystyle \forall x,y \in\Npq \quad x+y\in\Npq~.$
\end{pty}

\subsubsection{Different definition of rational bases}
\label{s.fk}

In this article, we use the original definition of rational bases \cite{AkiyEtAl08}.
In \cite{FrouKlou12}, the author define them
differently.  
Let us stress that the two definitions are not equivalent, although
they define two objects that are close.
Each has properties that the other has not, hence one object is
more convenient than the other depending on the context.

To make things precise, we give below the evaluation of a
word~$u=a_{k-1}\cdots a_0$ in~$\Aps$, as defined in \cite{FrouKlou12}.
\begin{align*}
  \valfk(u)&=\valfk(a_{k-1}\cdots a_0) =
  \sum^{k-1}_{i=0} a_i\left( \pq \right)^i 
  \\\notag
  &= q \xmd \val(u)
\end{align*} 
With this definition, the (rather strange) factor $\frac{1}{q}$
from \requation{eval} does not appear.
The representation algorithm is modified as follows (the difference is
underlined).
  \begin{align*}
    \repfk(0) ={}& \epsilon
    \\[1mm]
    \forall m\in\N\setminus\{0\}\quad
    \repfk(m) ={}&  \repfk(n) \cdot  a
      \quad\quad\text{where }\left\{
      \begin{array}{l}
        n\in \N,~ a\in\Ap \\
        q\xmd m = p\xmd n +\underline{qa}
      \end{array}\right.
  \end{align*}
Finally, note that these changes propagate to other objects such as the one below.
\begin{gather*}
  \Npqfk \overset{\text{def}}{=} \valfk{\Aps}  = q\Npq \subsetneq \Npq\\
  \Lpqfk \overset{\text{def}}{=} \repfk(\N) = \rep\left(\frac{1}{q}\N\right) = \Lpq\cdot \Ap \supsetneq \Lpq
\end{gather*}

The main statements of this article (that is, \rtheorem{pq-defi<->pq-reco},
\rproposition{npqrec.orde}, \rproposition*{pqrec.peri}, \rproposition*{insep},
and their corollaries) also hold, \emph{mutadis mutandis}, for this alternative
notion of rational base numeration systems.

\section{Characterisation of \protect\pqrec sets}

The purpose of section~\thesection{} is to show our main result.
We restate it after a few definitions.
\begin{defi}\hfill
  \begin{subthm}
    \item We let~$\Vpq$ denote the function~${\Npq\rightarrow\Npq}$
      that maps~$0$ to~$0$, and every positive~$x$ to~$\left(\frac{p}{q}\right)^k$,
      where~$k$ is the largest exponent such that~$x\left(\frac{p}{q}\right)^{-k}$ belongs to~$\Npq$.
    \item Let~$d$ be an integer and~$S$ be a subset of~$\Npqd$.
    We say that~$S$ is \pqdef if it is defined by some formula in $\logic$.
  \end{subthm}
\end{defi}
\begin{falsetheorem}{\rtheorem*{pq-defi<->pq-reco}}
  A set is~$\pq$-recognisable if and only if it is~$\frac{p}{q}$-definable.
\end{falsetheorem}

The forward direction of \rtheorem{pq-defi<->pq-reco}
is shown in \rsection{pq-reco->pq-defi} (\rproposition{pq-reco->pq-defi}) and
the backward direction in \rsection{pq-defi->pq-reco}
(\rproposition{pq-defi->pq-reco}).
First, we show some preliminary results in \rsection{length-preorder}.

\newcommand{\ALL}[1]{\textsf{all}_{#1}}
\newcommand{\zero}{\textsf{zero}}
\newcommand{\one}{\textsf{one}}
\newcommand{\norm}{\textsf{smallest-mask}}
\newcommand{\reduce}{\textsf{mask}}

\subsection{Constants and the length preorder are \protect\pqdef}
\lsection{length-preorder}

\begin{pty}\lproperty{vpq}
  Let~$x$ be a number in~$\Npq$ such that~$x\neq0$,
  and let~$i$ be a nonnegative integer.
  The following are equivalent.
  \begin{subthm}
    \item There exists a letter~$a\in\Ap$, $a\neq0$, such that~$a\xmd0^i$ is a
    suffix of~$\rep(x)$.
    \item It holds~$\Vpq(x) = \pqp^i =\val(q\xmd 0^{i})$.
  \end{subthm}
\end{pty}
\begin{proof}
  We denote the representation of~$x$ by~$\rep(x)=u\xmd a\xmd 0^{i}$
  with~$a\neq 0$.
  From~$\requation*{eval-recu}$, it holds~$\val(u\xmd a)=x\xmd\pqp^{-i}$; this
  number thus belongs to~$\Npq$.
  On the other hand, if~$x\xmd\pqp^{-(i\po)}$ belonged in~$\Npq$, it would be
  represented by some word~$w$, and then~$w\xmd 0^{i+1}$ would be an expansion
  of~$x$, a contradiction to the uniqueness of representation
  (\rtheorem{repr-uniq}).
  Hence, $\Vpq(x)= \pqp^{i}$.
  The other direction is similar.
\end{proof}

\begin{lem}\llemma{cons-pq-defi}
  Every constant~$c$ in~$\Npq$ is \pqdef.
\end{lem}
\begin{proof}
   First, let us show that there is a formula that realises 
   the natural order relation over~$\Vpq(\Npq)$, that is over the image of
   $\Vpq$, using the addition relation.
  \begin{claim}\lclaim{cons-pq-defi:1}
    Let~$x,y$ be two numbers in~$\Vpq(\Npq)$.
    There exists $z\in\Npq$ such that~$x=y+z$ if and only if~$x\geq y$.
  \end{claim}
  \begin{proofwithbar}\hfill
    \begin{itemize}
      \item The forward direction is immediate.
      \item Backward direction. The case~$x=0$ and the case~$y=0$ are trivial.
    If~$x\neq0$ and~$y\neq 0$ the representations of~$x$ and~$y$ are
    respectively written as~$\rep(x)=q0^i$ and~$\rep(y)=q0^j$ for some
    nonnegative integers~$i$ and~$j$.
    Hypothesis implies that~$i\geq j$ and it may be verified 
    that~$z=\val((p-q)^{i-j}0^j)$ satifies~$x=y+z$.
    \end{itemize}
  \end{proofwithbar}

  The number~$0$ is the neutral element for addition, hence the set \{0\} is
  defined by the following formula.
  \begin{equation*}
    \zero(x) \defeq\quad \forall y\quad x+y\mathbin{=}y\\
  \end{equation*}
  \rclaim{cons-pq-defi:1} allows us to define the set~$\{1\}$ with
  the following formula.
  \begin{equation*}
    \one(x) \defeq\quad \Vpq(x)\mathbin= x \relwedge \forall y~~\big(\,\Vpq(y)\mathbin=y \logimplies (\zero(y) \relvee \exists z~~x+z\mathbin{=}y\,\big))
  \end{equation*}
  Then, for every other number~$z$ in~$\Npq$, written as~$z=\frac{n}{m}$,
  the set~$\{z\}$ is \pqdef by the following formula.
  \begin{equation*}
  \textsf{constant}_{z}(x)\defeq\quad\quad
    \exists y~~\one(y) \relwedge \underbrace{\rule[-.75ex]{0pt}{1pt}y+y+\cdots+y}_{n\text{ times}}  =\underbrace{\rule[-.75ex]{0pt}{1pt}x+x+\cdots+x}_{m\text{ times}}
  \end{equation*}
  (Note that the formula $\textsf{constant}_{z}$
  is well formed for any~$z\in\Q$. However, if~$z$ does not belong to $\Npq$,
  this formula defines the empty set.)
\end{proof}

Since all constants are \pqdef, in the following, we use in formulas, directly
numbers as terms.
For instance we will write~~$\phi(x,y)\defeq x+\frac{p}{q} = y$~ instead
of~~$\phi(x,y)\defeq \exists z~ \textsf{constant}_{\frac{p}{q}}(z)\relwedge x+z = y$.

\begin{defi}
    We denote the length preorder over~$\Npq$ by~$\lord$ :
      for every~$x,y\in\Npq$, we write $x\lord y$ ~if~~$\wlen{\rep(x)}\leq\wlen{\rep(y)}$.
\end{defi}

The remainder of this subsection~\thesubsection{} is dedicated to showing
that relation~$\lord$ is definable in~$\logic$ (\rproposition{lord-pqrec}).
The general idea is that 
\begin{enuminline}
\item the natural order is \pqdef if it is restricted to the set of the numbers
whose representations use only the digit~$(p\mo)$, 
\item the function that maps a number~$x\in\Npq$ to~$(p\mo)^{\wlen{\rep(x)}}$ is
also~\pqdef.
\end{enuminline}
\begin{defi}
  Let us define a formula~$\ALL{a}(x)$, for each letter~$a\in \Ap, a\neq 0$.
  Equation~\eqref{eq.def-allpq} defines the case~$a=(p-q)$ and
  Equation~\eqref{eq.def-allnpq} the other cases.
\begin{align}\label{eq.def-allpq}
  \ALL{p-q}(x)\defeq{}& \exists z\quad \Vpq(z)\mathbin=z \relwedge x \mathop+ 1 \mathbin= z
  \\
  \label{eq.def-allnpq}
  \ALL{a}(x) \defeq{}& \exists y\quad \underbrace{\rule[-.75ex]{0pt}{1pt}x + x + \cdots + x}_{p-q\text{ times}} = \underbrace{\rule[-.75ex]{0pt}{1pt}y + y + \cdots + y}_{a\text{ times}} \quad{\wedge}\quad \ALL{p-q}(y)
\end{align}
\end{defi}
  
\begin{lem}
  For each~$a\in\Ap$, $a\neq0$, formula $\ALL{a}(x)$ defines the
  set~$\setq{x\in\Npq}{\rep(x)\in a^*}$. 
\end{lem}
\begin{proof}\hfill
  \begin{enumerate}
    \item Case~$a=p-q$. 
    First, note that Equation \eqref{eq.xxxx}, below, is a direct consequence
    of how carry is propagated in base~$\frac{p}{q}$.
    See the definition of the additionner (Equation~\eqref{eq.addi});
    and recall that the number~$1$ is represented by the one-letter word~$q$ in
    base~$\frac{p}{q}$.
    \begin{equation}\label{eq.xxxx}
      \forall x\in \Npq,~ n\in\mathbb{N} \quad \rep(x) = (p-q)^n \implies \rep(x+1) = q0^{n}
    \end{equation}

    \begin{itemize}
      \item Forward direction. Let~$x$ be a number in~$\Npq$ such that~$\ALL{p-q}(x)$ holds.
      Equation~\eqref{eq.def-allpq} implies that the number~$z=x+1$ is equal
      to~$(\frac{p}{q})^i$, for some nonnegative integer~$i$, hence written
      ${\rep(z)=q\xmd 0^i}$.
     From~\eqref{eq.xxxx}, the number~$x'=\val((p-q)^i)$ is such that
     ${x'+1=z}$, hence $x=x'$. 
     Since $\frac{p}{q}$-representation is unique up to leading zeroes,  it holds~$\rep(x)=(p-q)^{i}$.

    \item Backward direction is a direct consequence of \eqref{eq.xxxx}.
  \end{itemize}
  \smallskip
    \item Case~$a\neq p-q$ (and $a\neq 0$).
      Let~$x$ be a number in~$\Npq$ and~$n$ be a non-negative integer.
      The following equivalences conclude the proof.
      \begin{multline}
        \rep(x)= a^n 
        \iff x = \val(a^n) 
        \iff \frac{p-q}{a}\xmd x = \val((p-q)^n) 
        \\\iff \rep\left(\frac{p-q}{a}\xmd x\right) = (p-q)^n
      \end{multline}
      (Note that in \eqref{eq.def-allnpq}, the equality already implies that y is necessarily equal to~$\frac{p-q}{a}x$, if this number indeed belongs to $\Npq$.)\qedhere
  \end{enumerate}
\end{proof}

\noindent We now define a formula that enables to mask the actual value of a
number, but keeps the length of its representation.
Effectively, it replaces each digit by~$p\mo$ (that is, by the greatest digit).
This is done in two steps: $\reduce(x,z)$ is true if there is a word~$w$
such that~$\val(w)=x$ and $\val(w')=z$, where~$w'$ is the word resulting
from replacing every digit in~$w$ by $p\mo$; note that~$w$ may contain
leading $0$'s.
We say that~$z$ \emph{masks}~$x$ if $\reduce(x,z)$ is true.
Then, $\norm(x,y)$ is true if~$y$ is the smallest mask of~$x$, that is, if it
is masked by every mask of~$x$.
  \begin{gather}
    \reduce(x,z) := \ALL{p\mo}(z) \relwedge \exists y~~x+y\mathbin=z 
    \label{eq.def-reduce}
    \\
    \label{eq.def-norm}
    \norm(x,y) :=\reduce(x,y) \relwedge \forall z  \quad\reduce(x,z) \rightarrow \reduce(y,z)
  \end{gather}
Now, let us show that~$\reduce$ has the intended behavior.

\begin{lem}
  Let~$x$ and~$y$ be two numbers in~$\Npq$.
  \begin{subthm}
    \item \llemma{reduce}
    It holds $\reduce(x,z)$ if and only if~$\rep(z)$ belongs to~$(p\mo)^*$ and
    ${\wlen{\rep(z)}\geq \wlen{\rep(x)}}$.
    \item \llemma{norm}
    It holds $\norm(x,y)$ if and only if it holds~$y=\val((p\mo)^n)$, with ${n= \wlen{\rep(x)}}$.
  \end{subthm}
\end{lem}

\begin{proof}\hfill
  \begin{subthm}
  \item We write~$u=\rep(x)$ and~$n=\wlen{u}$.
  
  \begin{itemize}
    \item Forward direction.
    Obviously, $\rep(z)$ is equal to~$(p\mo)^m$ for some~$m$; it remains to show
    that~$m\geq n$. 
    The definition of the additioner (Equation \eqref{eq.addi})
    implies that no transition coming in state~$0$ outputs the digit~$0$, except
    the~$(0,0,0)$ loop.  
    In other words, the representation of a sum is never shorter than the
    representation of one of the operands.

    \item Backward direction.
    Let~$m$ be the nonnegative integer such that~$(p\mo)^m = \rep(z)$,
    hence by hypothesis we have~$m\geq n$.
    Let~$a_{m\mo}\cdots a_1 a_0 = 0^{m-n} \rep(x)$.  For each integer~$i$, $0\leq
    i<m$, the number~$(p-1-a_i)$ belongs to~$\{0,1,\ldots,p\mo\}$, hence is a
    digit in~$A_p$.
    Hence, the word~$w=(p-1-a_{m\mo})\cdots(p-1-a_{1})(p-1-a_0)$
    belongs to~$(A_p)^*$
    and~$y=\val(w)$ belongs to~$\Npq$.
    It is clear that~$x+y=z$ since no carry will be risen during this addition;
    hence~$y$ is the witness that~$\reduce(x,z)$ holds. 
  \end{itemize}
  \item From Item (a), any number with representation~$(p\mo)^n$ may be
  reduced to any number with representation~$(p\mo)^m$ if~$n\geq m$.  
  The statement follows.\qedhere
  \end{subthm}
\end{proof}

Now, all is set to show that relation~$\lord$ is \pqdef. 

\begin{prop}\lproposition{lord-pqrec}
  The relation~$\lord$ is definable in~$\logic$.
\end{prop}
\begin{proof} Let us show that the following formula defines~$\lord$.
  \begin{equation*}
    \Phi(x,y) \defeq \exists x'~\exists y'\quad \norm(x,x') \relwedge \norm(y,y') \relwedge \reduce(x',y')
  \end{equation*}

  \begin{itemize}
    \item Forward direction. Let~$x,y\in\Npq$ be two numbers such that~$\Phi(x,y)$ is true.
  From \rlemma{norm}, the following two equations hold.
  \begin{gather*}
    x'= \val{(p\mo)^n} \quad\text{with }n=\wlen{\rep(x)} \\
    y'= \val{(p\mo)^m} \quad\text{with }m=\wlen{\rep(y)}
  \end{gather*}
  From \rlemma{reduce},  since $\reduce(x',y')$ is true, it holds~$n\leq m$,
  hence~$x\lord y$.

    \item Backward direction. Let~$x,y\in\Npq$ be two numbers such that~$x\lord y$.
  We let~$n$ and~$m$ denote the length of the representation of~$x$ and~$y$,
  respectively; hence~$n\leq m$.
  The numbers~$x'=\val((p-1)^n)$ and~$y'=\val((p-1)^m)$ are the witnesses
  that~$\Phi$ true.
  \qedhere
  \end{itemize}
\end{proof}

\subsection{Every \protect\pqrec set is \protect\pqdef}
\lsection{pq-reco->pq-defi}
Section \thesubsection{} is dedicated to the proof of
\rproposition{pq-reco->pq-defi}.
It is adapted from the proof of \rtheorem{buch-bruy} given in \cite{BruyEtAl94}.
The main idea is to code the run of a $d$-state automaton, that is a sequence
of elements from~$\{0,1,\ldots,d\mo\}$, by a $d$-tuple of 
numbers in~$\Npqd[d]$.
The \pqreps of the numbers in the $d$-tuple contain only $0$'s and $1$'s, and
are the characteristic functions of the respective state.
For instance, with~$d=4$, the sequence
$(s_0,s_1,s_2,s_3,s_4,s_5,s_6)=(3,0,3,1,1,1,1)$ is coded by the $4$-tuple
$(x_0,x_1,x_2,x_3)\in\Npqd[4]$ defined by the following.
\begin{equation}
  (\rep(x_0),\rep(x_1),\rep(x_2),\rep(x_3))=(10,1111000,\epsilon,101)
\end{equation}
Indeed, the digit at position~$0$ in~$\rep(x_3)$ is~$1$ because~$s_0=3$; the
digit at position~$1$ in~$\rep(x_0)$ is~$1$ because~$s_1=0$, etc.

\medskip

\rdefinition{extr-pred}, below, gives a few relations and functions that will
be useful in the proof and \rlemma{extr-pred} shows that they are \pqdef.
In particular, it gives the formula that computes the digit at a given position
in the representation of a given number.
First, let us clarify what we mean by \emph{position} in a representation.
Let~$x$ be a number in~$\Npq$, the representation of which we
write~$\rep(x)=a_{n-1}\cdots a_1\xmd a_0$.
For every nonnegative integer~$i$, the \emph{digit at position~$i$
in~$\rep(x)$} refers to~$a_i$ if~$i<n$, and to~$0$ otherwise.

\begin{defi}\ldefinition{extr-pred}\hfill
  \begin{subthm}
    \item Let~$\Vpqp$ be the function~${\Npq\rightarrow\Npq}$ that maps~$0$ to~$0$
      and every positive~$x$ to~$\val(1\xmd 0^i)$, where~$i$ is the
      integer such that~$\rep(x)=u\xmd a\xmd 0^{i}$, with~$a\neq 0$.
    \item We write~$\bpq(x)$ if~$x$ if the
          representation of~$x$ belongs to the language~$1\xmd 0^*$.
    \item Similarly to~$x\lord y$, we write $x\lord< y$ if~$\wlen{\rep(x)}<\wlen{\rep(y)}$.
    \item For each digit~$a\in\Ap$,  we let~$\ast_{a}(y,z)$ denote the
    multiplication-by-a-digit relation: it
    holds~$\ast_{a}(y,z)$ and~$z=a\times y$.
    \item For each digit~$a\in\Ap$, let~$\digit_{a}(x,y)$ be the relation that
    holds if~$\bpq(y)$ and if~$a$ is the digit at position~$i$ in~$\rep(x)$,
    where~$i$ is the position of the unique digit~$1$ in~$\rep(y)$.
  \end{subthm}
\end{defi}

\begin{lem}\llemma{extr-pred}
  Functions and relations from \rdefinition{extr-pred} are all \pqdef.
\end{lem}
\begin{proof}\hfill
\newcommand{\vmstrut}{\rule[-1.5ex]{0pt}{3.5ex}}%
\newcounter{vmtmp}\stepcounter{vmtmp}%
\renewcommand{\thevmtmp}{\alph{vmtmp}}%
\newcommand{\vmtag}{\tag{\stepcounter{vmtmp}\thevmtmp}}%
\makeatletter\tagsleft@true\let\veqno\@@leqno\makeatother%

\noindent(\thevmtmp)
\hfill$\displaystyle
  \Vpqp(x) = z \quad{{}\defeq} \quad \Vpq(x)=\underbrace{\rule[-.75ex]{0pt}{1ex}z+\cdots+z}_{q\text{ times}}
$\hfill\hskip0pt\\
If~$x=0$, the above formula is obviously correct.
Otherwise, from \rproperty{vpq}, it holds ${\Vpq(x)= \pqp^{i}}$ and~$z=\frac{1}{q}\pqp^i$.
\requation{eval} yields that~$z=\val(1\xmd 0^i)$.
\begin{align*}
  \vmtag \bpq(x) \defeq&\quad \Vpqp(x)\mathbin{=}x \relwedge \neg (x \mathbin= 0) \\
  \vmtag
          x\mathbin{\lord<}y \defeq
    &\quad
            x\mathbin{\lord\leq} y \relwedge \neg (y\mathbin{\lord\leq} x) \\
  \vmtag
   \ast_{a}(y,z) \defeq&\quad   z \mathbin{=} \underbrace{y+\cdots+y}_{a\text{ times}} \hspace*{-2cm}
\end{align*}
It is a routine to show that the formula defined by \enumstyle{b},
\enumstyle{c} and~\enumstyle{d} are correct.
\begin{multline*}
  \vmtag  \digit_{a}(x,y) \defeq \\
  \smash{\underbrace{\bpq(y)}_{\enumstyle{i}}} \relwedge \exists \ell~\exists m~\exists r
  \quad\Bigg(
      \smash{\overbrace{\vmstrut x\mathbin{=}\ell+m+r}^{\enumstyle{ii}}}
            \relwedge \smash{\overbrace{\vmstrut{\ast_{a}}(y,m)}^{\enumstyle{iii}}} \\
            {\relwedge{}} \underbrace{\Big(\vmstrut \ell\mathbin{=}0 \relvee y\mathbin{\lord<}\Vpqp(\ell)\Big)}_{\enumstyle{iv}}
            \relwedge \underbrace{\vmstrut r\mathbin{\lord<} y}_{\enumstyle{v}}
\Bigg)
\end{multline*}
\indent Assume that~$\digit_{a}(x,y)$, as defined by \enumstyle{e} is true.
  Term~\enumstyle{i} implies that there is an
  integer~$i$ such that~$\rep(y)=10^i$.
  The variable~$\ell$,~$m$ and~$r$ will contain the left, middle and right parts
  of~$x$ where the split is done at position~$i$ of the representation of~$x$.
  Term~\enumstyle{iii} ensures that~$\val(a\xmd 0^i)=m$.
  Term~\enumstyle{iv} ensures that
  there is a word~$u\in\Aps$ such that~$\val(u\xmd 0^{i+1})=\ell$.
  Term~\enumstyle{v} ensures that~$\rep(r)=v$ for some word~$v$ such that~$\wlen{v}\leq i$.
  Then,~\enumstyle{ii} implies that
  \begin{equation*}
    x=\val(u\xmd 0^{i+1})+\val(a\xmd 0^i)+\val(v) = \val(u\xmd a\xmd 0^jv)
    \quad\text{where}\quad j=i-\wlen{v}~.
  \end{equation*}
  Hence, the digit at position~$i$ in~$\rep(x)$ is~$a$.
  Conversely, let~$x$ be any number in~$\Npq$ and~$y$ be a number in~$\Npq$
  such that~$y=\frac{1}{q}\big(\pq\big)^i$ for some integer~$i$.
  For~$k$ great enough, we factorise~$0^k\rep(x)$ as~$u \xmd a\xmd v$  where~$u$ and~$v$ are two words in~$\Aps$
  such that~$\wlen{v}=i$, and~$a$ is a letter in~$\Ap$.
  The values $\ell=\val(u)$, $m=\val(a\xmd 0^i)$
  and~$r=\val(v)$ will be the witnesses that~$\digit_{a}(x,y)$, as defined by
  \enumstyle{e} is true.
\end{proof}

Now, let us show the forward direction of~\rtheorem{pq-defi<->pq-reco},
restated below.

\begin{prop}\lproposition{pq-reco->pq-defi}
  Every \pqrec set is~$\pq$-definable.
\end{prop}
\begin{proof}
  Let~$S$ be a subset of~$\Npqd$ that is \pqrec, hence realised
  by a~$d$-tape automaton~$\Ac$.
  We let~$m$ denote the number of states in~$\Ac$.
  Without loss of generality, we assume that~$\Ac$ is right-to-left, complete,
  that the state-set is~$\{0,1,\ldots, m-1\}$ and that the initial state is~$0$.
  We moreover denote by~$F$ the set of final states and by~$\delta$ the
  transition function. 
  In summary, $\Ac = \langle \Apd,\,\{0,1,\ldots, m-1\}, 0, \delta, F
  \rangle$~.

  \smallskip

  Let~$\vect{x}=(x_0,x_1,\ldots,x_{d\mo})$ be a~$d$-tuple of numbers.
  The formulas will describe the run of the word~$\vect{u}=\rep(\vect{x})$.
  First, we introduce a variable~$k$ equal to~$\val(1 0^{K})$
  where~$K$ is the length of~$\vect{u}$.
  Hence,~$K$ is equal to the max of the
  lengths~$\wlen{\rep(x_0)}, \ldots,\wlen{\rep(x_{d\mo})}$.
  Second, we introduce one variable per state of~$\Ac$, denoted
  by~$s_0,s_1,\ldots,s_{m\mo}$.
  The proof consists in proving that, for every integer~$n$, $0\leq n \leq
  \wlen{\vect{u}}$,
  \begin{multline}\lequation{pq-defi<-pq-reco-pre}
    \forall i,~0\mathbin{\leq} i \mathbin{<} m,\\
    \text{the digit at position~$n$ in~$\rep(s_i)$ is}\quad\left\{
    \begin{array}{l@{}l}
    1\quad&\text{if the run of~$\suff{n}(\vect{u})$ reaches state~$i$}\\
    0 &\text{otherwise}
    \end{array}
    \right. \\
    \text{where $\suff{n}(\vect{u})$ is the suffix of~$\vect{u}$ of length~$n$.}
  \end{multline}

  \smallskip

  \noindent\requation{pq-defi<-pq-reco-L}, below, gives the formula~$\Lambda(\fcdots)$ that defines~$k$.
  \begin{multline}\lequation{pq-defi<-pq-reco-L}
    \Lambda(x_0,x_1,\ldots,x_{d\mo},k) {{}\defeq}
    \\
    \bpq(k)\relwedge{} \left(\bigwedge_{i=0}^{d-1}  x_i\mathbin{\lord<} k \right)
    \relwedge{} \left(\forall j \quad \big(\bpq(j) \relwedge j \mathbin{\lord<} k\big)
    \logimplies \bigvee_{i=0}^{d-1}j\mathbin{\lord\leq} x_i \right)
  \end{multline}

  \begin{claim}
    The formula~$\Lambda(x_0,x_1,\ldots,x_{d\mo},k)$ is true 
    if and only if
    $k=\frac{1}{q}\left(\frac{p}{q}\right)^K$,
    where ${K=\textsf{max}~\big\{\wlen{\rep(x_i)}\big\}_{0\leq i< d}}$.
  \end{claim}
  \begin{proofwithbar}
  It is a routine to establish the following equivalence.
  \begin{itemize}
    \item The formula $\bpq(k)$ is true if and only ${k=\frac{1}{q}\left(\frac{p}{q}\right)^{K'}}$, where ${K'=\left(\wlen{\rep(k)}-1\right)}$.
    \item The formula~$\bigwedge_{i=0}^{d-1}  (x_i\mathbin{\lord<} k)$ is true
    if and only if it holds~$\wlen{\rep(k)}>\wlen{\rep{x_i}}$, for each $i$, $0\leq i < d$.
    \item The formula
    $
      \left(\forall j \quad \big(\bpq(j) \relwedge j \mathbin{\lord<} k\big)
      \logimplies \bigvee_{i=0}^{d-1}(j\mathbin{\lord\leq} x_i) \right)$
    is true if and only if, for each~${J<K}$, there exists~$i$ such that~$\wlen{\rep(\frac{1}{q}(\frac{p}{q})^J)}\leq\wlen{\rep(x_i)}$.
  \end{itemize}%
  Claim~\theclaim{} immediately follows.
  \end{proofwithbar}
  
  \medskip

  \noindent\requation{pq-defi<-pq-reco-I}, below, defines~$\xi(\fcdots)$ expressing
  that~$0$ is the initial state.
  %
  \begin{equation}
  \lequation{pq-defi<-pq-reco-I}
    \xi(s_0,s_1,\ldots,s_{m\mo})\defeq\quad
    \digit_1\left(s_0,\frac{1}{q}\right) \relwedge\bigwedge_{i=1}^{m-1} \digit_0\left(s_i,\frac{1}{q}\right)
  \end{equation}

  \begin{claim}
    The formula $\xi(s_0,\ldots,s_{m\mo})$
  is true if and only if
      \requation{pq-defi<-pq-reco-pre} holds for~$n=0$.
  \end{claim}
  \begin{proofwithbar}
  Recall that~$\frac{1}{q}=\val(1)$.
  Hence $\xi(s_0,s_1,\ldots,s_{m\mo})$ is true if and only if both
  \begin{itemize}
    \item the digit at position $0$ in~$\rep(s_0)$ is $1$, and
    \item for each~$i$,~$0<i<m$, the digit at position $0$ in~$\rep(s_i)$ is 0.
  \end{itemize}
  This concludes the proof of Claim \theclaim{} since~$0$ is the initial state.
  \end{proofwithbar}

  \medskip
  
  \noindent Provided that~$k$ is of the form~$k=\frac{1}{q}\left(\frac{p}{q}\right)^K$, \requation{pq-defi<-pq-reco-F}, below, defines~$\Phi(\fcdots)$ which
  is true if the run of $\suff{K}(\vect{u})$ reaches a
  final state.
  (Recall that~$F$ is the set of final states.)
  \begin{equation}\lequation{pq-defi<-pq-reco-F}
    \Phi(s_0,s_1,\ldots,s_{m-1},k) {{}\defeq}\quad
      \bigvee_{i\in F} \digit_{1}(s_i,k)
  \end{equation}
  
  The following claim immediately follows.
  \begin{claim}
    The formula~$\left(\Lambda(x_0,x_1,\ldots,x_{d\mo},k)\relwedge \Phi(s_0,s_1,\ldots,s_{m-1},k)\right)$ is true if and only if the following two conditions hold.
    \begin{subthm}
      \item $k=\frac{1}{q}\left(\frac{p}{q}\right)^K$,
        where $K=\textsf{max}~\big\{\wlen{\rep(x_i)}\big\}_{0\leq i< d}$
      \item There exists an integer~$i\in F$, such that the digit at
        position~$K$ in~$s_i$ is 1.
    \end{subthm}
  \end{claim}

  \medskip

  \requation{pq-defi<-pq-reco-D}, below,  defines~$\Delta(\fcdots)$ that will
  ensure that \requation{pq-defi<-pq-reco-pre} is inductively satisfied.
  (Recall that~$\delta$ is the transition function of~$\Ac$).  \begin{multline}
    \Delta(x_0,\cdots,x_{d\mo}, s_0,\cdots,s_{m\mo},j) {{}\defeq}
    \\
    \exists h \quad \overbrace{\rule{0pt}{2ex}j+\cdots+j}^{p\text{ times}} = \overbrace{\rule{0pt}{2ex}h+\cdots+h}^{q\text{ times}}
    \quad\quad\quad\quad\quad\quad\quad\quad\quad\quad\quad\quad\quad\quad\quad\quad
    \\
    \bigwedge_{\substack{\vect{a}\,=\,(a_0,\cdots,a_{d\mo})\in\Apd\\0\,\leq\,i\,<\,m}}
      \Bigg(\Big(\digit_{1}(s_i,j) \relwedge \bigwedge_{\ell=0}^{d-1} \digit_{a_\ell}(x_\ell,j)\Big) \\
%
%
    \lequation{pq-defi<-pq-reco-D} \hspace*{.5cm}\mathbin{\logimplies}
      \Big(\digit_{1}(s_{\delta(i,\vect{a})},h) \relwedge
      \bigwedge_{\substack{0\,\leq\,g\,<\,m \\g\,\neq\,\delta(i,\vect{a})}} \digit_{0}(s_g,h) \Big)\Bigg)
  \end{multline}
  \begin{claim}
    Let~$J$ be a nonnegative integer. We assume that
    \requation{pq-defi<-pq-reco-pre} holds for ${n=J}$, and we
    write~$j=\frac{1}{q}(\frac{p}{q})^J$.
    The formula $\Delta(x_0,\cdots,x_{d\mo}, s_0,\cdots,s_{m\mo},j)$ is true
    if and only if~\requation{pq-defi<-pq-reco-pre}
    holds for~$n=(J+1)$.
  \end{claim}
  \begin{proofwithbar}
    Since~\requation{pq-defi<-pq-reco-pre} holds for~$n=J$,
    there is exactly one integer~$i$ in $\{0,\ldots,m\mo\}$ such that
    $\digit_{1}(s_i,j)$ is true;
    and~$i$ is the state reached by the run of~$\suff{J}(u)$, the suffix of
    length~$J$ of~$u$.
    Moreover for each~$\ell$, $0\leq\ell<d$, we let~$b_\ell$ denote
    the digit such that~$\digit_{b_\ell}(x_\ell,j)$ is true.
    Finally, we write~$i'=\delta(i,(b_0,\ldots,b_{d\mo}))$,
    or, in other words, $i'$ is the state reached by the run
    of~$\suff{(J\po)}(u)$.
    
    In this context, $\Delta(x_0,\cdots,x_{d\mo}, s_0,\cdots,s_{m\mo},j)$ is
    equivalent to the following.
    \begin{multline*}
    \exists h \quad \underbrace{\rule[-.75ex]{0pt}{0pt}j+\cdots+j}_{p\text{ times}} = \underbrace{\rule[-.75ex]{0pt}{0pt}h+\cdots+h}_{q\text{ times}}
    \relwedge
      \Big(\digit_{1}(s_{i'},h) \relwedge
      \bigwedge_{\substack{0\,\leq\,g\,<\,m \\g\,\neq\,i'}} \digit_{0}(s_g,h) \Big)
  \end{multline*}
  The first conjunct is true if and only if~$h=\val(1 0^{J+1})$,
  hence the whole formula is true if and only if both
    \begin{itemize}
      \item the digit at position~$J+1$ in~$\rep(s_{i' })$ is $1$; and
      \item the digit at position~$J+1$ in~$\rep(s_{g})$ is $0$ if~$g\neq i'$.
    \end{itemize}
  This concludes the proof of Claim~\theclaim{}.
  \end{proofwithbar}

  \smallskip

  Finally, \requation{pq-defi<-pq-reco-tota} gives the
  formula~$\Omega(\fcdots)$ of the logic $\logic$ that defines the set~$S$.
  We use~$\vect{x}$ and~$\vect{s}$ has shorthand for~$(x_0,x_1,\ldots,x_{d\mo})$
  and~$(s_0,s_1,\ldots,s_{m\mo})$.
  \begin{multline}
    \lequation{pq-defi<-pq-reco-tota}
    \Omega(x_0,x_1,\ldots,x_{d\mo}) \defeq \\
    \exists k~\exists \vect{s}\quad
     \Lambda(\vect{x},k)
     \relwedge \xi(\vect{s})
     \relwedge \Phi(\vect{s},k)
     \relwedge 
     \forall j~\Big(  \big(\bpq(j)
    \relwedge j \mathbin{\lord<} k \big) \logimplies \Delta(\vect{x},\vect{s},j)\Big)
  \end{multline}
  It follows from the previous claims that~$\Omega(\fcdots)$ defines the
  set~$S$, concluding the proof of Proposition~\thethm{}.
\end{proof}

\subsection{Every \protect\pqdef set is \protect\pqrec}
\lsection{pq-defi->pq-reco}

The backward direction of \rtheorem{pq-defi<->pq-reco} is proved in a very
classical way.  We only give a sketch of the proof.
It amounts to showing that 1) each atomic set or relation of the logic is
realised by an automaton 2) the inductive constructs of first-order formulas
preserves \pqrecy.
For more details, see for instance \cite{BruyEtAl94}.

\begin{prop}\lproposition{pq-defi->pq-reco}
  Every~$\frac{p}{q}$-definable set is~$\pq$-recognisable.
\end{prop}
\begin{proof}[Sketch of proof]
  Let~$\Ac$ be a $d$-tape complete automaton
  that realises a $d$-free-variable formula~$\Phi$.
  %
\begin{enumerate}
    \item The set defined by the formula~$\neg\Phi(\vect{x})$ is realised by
    the automaton resulting from inverting final and non-final states in~$\Ac$.
    \item The set defined by the formula~$\exists x_i~~\Phi(\vect{x})$ is
    realised by the automaton resulting from erasing the~$i$-th tape
    from~$\Ac$.  
    (Note that the resulting automaton needs to be determinised and made
    padded.)
    \item The formula $\forall x_i~~\Phi(\vect{x})$ is equivalent to~$\neg\Big(
    \exists x_i~~\neg\Phi(\vect{x})\Big)$.
    \item Let~$S$ be the set of~$\Npqd[d+1]$ defined
      by~
      \begin{equation*}
        \Theta(x_0,x_1,\ldots,x_i,y,x_{i+1},\ldots,x_{d\mo}) \defeq\quad \Phi(\vect{x})
      \end{equation*}
      that is, where the "free" variable~$y$ is added without being used.
      $S$ is realised by~$\Cc$ defined as follows.
    $\Cc$ is a~$d+1$-tape automaton and it has the same state set, initial state and final-state set as~$\Ac$.
    The transition function of~$\Cc$ is defined by, for all appropriate~$a_i$'s,~$b$ and~$s$,
      \begin{equation}
        \hspace*{1cm}\delta_\Cc\big(s,(a_0,\ldots a_{i},b,a_{i+1},\ldots,a_{d\mo})\big) = \delta_\Ac\big(s,(a_0,\ldots, a_{d\mo})\big)~.
      \end{equation}
    \item It is a routine to build the automaton that realises the set~$S$ defined by
      \begin{equation*}
        \Pi(x_0,\ldots,x_{i\mo},x_k,x_{i+1}, \ldots,x_{k\mo},x_{i},x_{k\po}\ldots,x_{d\mo}) \defeq\quad\Phi(\vect{x})
      \end{equation*}
      that is, where the variables at positions~$i$ and~$k$ have swapped
      positions.
\end{enumerate}
Let $\Bc$ be a second $d$-tape complete automaton that realises a
$d$-free-variable formulas~$\Psi$. (Note that we may assume that~$\Psi$
and~$\Phi$ have the same free variables in the same order thanks to items
(d) and (e), above.)
\begin{subthm}[label=\textbullet]
    \item The set defined by~$\Phi(\vect{x}) \relwedge \Psi(\vect{x})$ is realised by the product automaton ${\Ac\times\Bc}$.
    \item The formula $\Phi(\vect{x}) \relvee
    \Psi(\vect{x})$ is equivalent to~$\neg\big(\neg\Phi(\vect{x}) \relwedge
    \neg\Psi(\vect{x})\big)$
\end{subthm}

\begin{figure*}
  \center
  \includegraphics[scale=\AutomatonScale,trim=0ex 0ex 0ex 0ex]{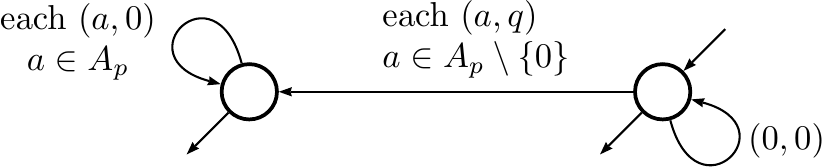}
  \caption{Right-to-left automaton that realises~$\Vpq$}
  \lfigure{wpq}
\end{figure*}

\medskip

  It remains to show that equality, addition, and the function~$\Vpq$
  are all \pqrec.
  Since, the~$\pq$-expansions of a number are equal up to leading~$0$'s,
  equality is obviously \pqrec.
  We previously recalled in \requation{addi}
  the definition of the automaton that realises addition in base~$\pq$.
  The function $\Vpq$ is realised by a simple automaton with two states,
  the definition of which is given as a prototype in~\rfigure{wpq}.
  %
\end{proof}

\section{Notable sets that are or are not \protect\pqrec}
\lsection{nota}

During the research leading to \rtheorem{pq-defi<->pq-reco}, it became apparent
that very little was known on the expressive power of automata with respect to
base~$\pq$.
After recalling a few known or obvious \pqrecy results, we show that the
natural order relation over~$\Npq$ is not \pqrec.
Then, we study generalisations of the modulo-$n$ operator in
Sections~\rsection*{modu-pqrec} and \rsection*{modu-not-pqrec}.
We show that it is \pqrec
if $n$ is coprime with~$q$, and that it is not if~$n$ is a multiple of~$q$.
These results are not immediate consequences of~\rtheorem{pq-defi<->pq-reco},
although it is sometimes useful in the proofs.

\subsection{A few known results}

\begin{prop} The following are~$\pq$-recognisable.
  \begin{subthm}
  \item \label{p.pqrec.addi}
  Addition: $\setq{(x,y,z)\in{\Npq}^3}{x+y=z}$.
  \item \label{p.pqrec.subt}
  Partial subtraction: $\setq{(x,y,z)\in{\Npq}^3}{x-y=z}$
  \item \label{p.pqrec.mult}
  Multiplication by a constant: $\setq{(y,z)\in{\Npq}^2}{r\xmd y=z}$, where~$r\in\Q$.
  \item \label{p.prec.inte}
  Interval bounded by constants:  $\setq{z\in{\Npq}}{r<z<s}$, where~$r$ and~$s$ belong to~$\R$.
  \end{subthm}
\end{prop}

Item \rproposition*{pqrec.addi} is given in \rsection{defi-rati-base}.
Item \rproposition*{pqrec.subt} and
\rproposition*{pqrec.mult} follows immediately from the fact that addition
is \pqrec (\rtheorem{addi}).
Item \rproposition*{prec.inte} follows from the fact that such an interval
contains a finite number of elements from~$\Npq$.
%


\begin{prop} The following are not~$\pq$-recognisable.
  \begin{subthm}
    \item \label{p.npqrec.inte.subset}
      Any infinite subset of~$\N$.
    \item \label{p.npqrec.fini-gene} Any finitely-generated additive submonoid of~$\Npq$, \ie of the
    form~$x_0\N + \cdots+x_n \N$, with~$x_0,\ldots,x_n\in\Npq$.
  \end{subthm}
\end{prop}

%
%
Item \rproposition*{npqrec.inte.subset} is a consequence of the
fact that~$\Lpq$ possess the \emph{Finite Left Iteration Property}, (\cf for
instance \cite{MarsSaka17b}).
Item \rproposition*{npqrec.fini-gene} is the main result of a
previous work of the author and Sakarovitch~\cite{MarsSaka13b}.
%

\subsection{Natural order}
\lsection{orde-not-pqrec}

The purpose of this section is to show that the order relation is not \pqrec
(\rproposition{npqrec.orde}).
First, we state an elementary property of~$\Npq$.

\begin{lem}
\llemma{vpq-ulti-all}
  For every integer~$k$,
  there is a bound~$m_k$ such that~$\frac{n}{q^k}$ belongs
  to~$\Npq$, for every integer~$n\geq m_k$.
\end{lem}
\begin{proof}
  The words~$10^{k-1}$ and~$1$ are evaluated to~$\frac{p^{k\mo}}{q^k}$
  and~$\frac{1}{q}$, respectively; hence, both numbers belong to~$\Npq$.
  Since~$\Npq$ is stable by addition, it 
  contains~$\frac{1}{q^k}(p^{k\mo}i+q^{k\mo}j)$, for every nonnegative
  integers~$i$ and~$j$.
  Since~$p$ and~$q$ are coprime, when~$j$ runs
  through~$\{0,1,\ldots,p^{k-1}\}$, then~$q^{k-1}j$ runs through all residue
  classes modulo~$p^{k-1}$.
  It follows that we may choose $m_k=(pq)^{k\mo}$.
\end{proof}

The bound~$m_k$ given is the proof is far from tight; a better, and seemingly
tight, bound is given by \cite[Lemme~5.15]{Mars16}.
In the following, we will use \rlemma{vpq-ulti-all} under a different form,
given below.

\begin{lem}\llemma{vpq-all-inte}
  For all real numbers~$\alpha$ and~$\beta$, such that ${0\leq\alpha<\beta}$,
  there exist two numbers~$x,y$ in~$\Npq$
  such that
  \begin{equation*}
    \alpha < x - y < \beta~.
  \end{equation*}
\end{lem}
\begin{proof}
  There exist two integers~$k$ and~$h$ such that~$\alpha<\frac{h}{q^k}<\beta$.
  Then, we set~$x=\frac{m_k+h}{q^k}$ and~$y=\frac{m_k}{q^k}$, where~$m_k$ is the
  bound given by \rlemma{vpq-ulti-all}.
\end{proof}


\begin{prop}\lproposition{npqrec.orde}
  The set~$\setq{\left(x,y\right)\in{\Npq}^2}{x<y}$ is not \pqrec.
\end{prop}
\begin{proof}[Proof]
  Let~$L=\setq{(v,w)\in\Apds[2]}{\val(v)<\val(w)}$.
  The proof amounts to showing that~$L$ is not a regular language.
  For the sake of contradiction, we assume that~$L$ is accepted by a~$n$-state
  automaton~$\Ac$ that we assume left-to-right, complete and minimal.
  We write~$\ell=\wlen{\rep(n)}$.
  For every~$i$, $0\leq i \leq n$, \rlemma{vpq-all-inte}, applied
  to~$\alpha=i\xmd\big(\pq\big)^{-\ell}$ and~$\beta=(i+1)\big(\pq\big)^{-\ell}$,
  yields two numbers~$x,y$ in~$\Npq$ and we write~$(v_i,w_i)=\rep((y,x))$.
  We have just defined~$n\po$ words in~$\Apds[2]$ such that
  \begin{equation}\lequation{order-i}
    \forall i,~0\mathbin{\leq} i\mathbin{\leq} n, \quad
    i < \left(\val(w_i\xmd 0^\ell)-\val(v_i\xmd 0^\ell)\right) < i+1~.
  \end{equation}

  %
  Let~$i$ and~$j$ be integers such that~$0\leq i,j\leq n$.
  We write~$\vect{u}_j=\big(0^{\ell-k}\rep(j),0^{\ell}\big)$,
  with~$k=\wlen{\rep(j)}$.
  %
  It follows that the two components of~$(v_i,w_i)\cdot \vect{u}_j$
  are evaluated to
  ${\big(\val(v_i\xmd 0^\ell)+j\big)}$ and~${\val(w_i\xmd 0^\ell)}$, respectively.
  Hence, from \requation*{order-i},~$(v_i,w_i)\cdot \vect{u}_j$ belongs to~$L$
  if and only if~$i\geq j$.
  Let~$M=\set{\vect{u}_1,\ldots,\vect{u}_n}$.
  The next equation sums up the previous paragraph.
  %
  \begin{equation*}
    \forall i,~0\mathbin\leq i\mathbin\leq n,\quad
    \big((v_i,w_i)^{\mo}L\big) \cap M ~=~ \set{\vect{u}_1,\ldots,\vect{u}_i}
  \end{equation*}
  Thus, the languages~$\left(\rule{0ex}{2.25ex}(v_i,w_i)^{\mo}L\right)_{0\leq
  i\leq n}$ are pairwise distinct,
  hence ${\card{Q}\geq n+1}$, where
  $(Q=\left\{(v,w)^{\mo}L~\middle|~(v,w)\in\Apds[2]\right\})$.
  On the other hand, since~$\Ac$ is minimal and complete, its state set is
  precisely~$Q$, hence~$\card{Q}=n$, a contradiction.
\end{proof}

\subsection{Equivalence modulo~$n$, where~$n$ is an integer coprime with~$q$}
\lsection{modu-pqrec}

The function~$x\mapsto (x\mod{n})$ classically maps integers to elements of~$\Z/n\Z$.
If~$n$ is coprime with~$q$, we may extend this function to~$\Npq\rightarrow\Z/n\Z$
as follows.
We let~$q^{\mo}$ denote the element of~$\Z/n\Z$ such that~$qq^{\mo}\mod{n} =
1$.
Let~$x$ be a number in~$\Npq$.
There exist integers~$m,k$ such that~$x=\frac{m}{q^k}$, and
we set
\begin{equation}\lequation{mod}
  x\mod{n}=\left(\frac{m}{q^k}\right) \mod{n}=m (q^{\mo})^k \mod{n}\quad.
\end{equation}
It may be verified that this generalised modulo-$n$ operator
still distributes over addition and multiplication.
It is then quite elementary to show that the computation of equivalence
classes modulo~$n$ is \pqrec.
\begin{prop}\lproposition{pqrec.peri}
  Let~$n$ be an integer coprime with~$q$, and let $R\subseteq Z/nZ$ be a set of
  remainders modulo~$n$.
  Then, the set~$P_{n,R}=\setq{ z\in\Npq}{(z\mod{n})\in R}$ is \pqrec.
\end{prop}
\begin{proof}[Sketch of proof]
  Once again,~$q^{\mo}$ denotes the element of~$\Z/n\Z$ such that~$(qq^{\mo})$
  is equivalent to~$1$ modulo~$n$.
  Let~$\Ac$ be the left-to-right automaton defined as follows.
  \begin{equation}\lequation{modu-n}
    \Ac = \langle\, \Ap,\, \Z/n\Z,\, \delta_\Ac,\, 0,\, R \,\rangle\quad
    \quad\quad\text{where}\quad\delta_\Ac(s,a) = q^{\mo}(s\xmd p\xmd + a) \mod{n}
  \end{equation}
  Using Equation~\requation*{eval-recu} and~\requation*{modu-n}, one may show
  with an induction over the length of~$u$ that, for every word~$u$ in~$\Aps$
  the run of~$u$ in~$\Ac$ exists and reaches the
  state~$\big(\val(u)\mod{n}\big)$.
\end{proof}
\begin{cor}\lcorollary{pqrec.modu}
  Let~$n$ be an integer coprime with~$q$.
  The following relation is \pqrec.
  \begin{equation}
    \setq{(x,y)\in \Npqd[2]}{x\mod{n}\mathbin{=}y\mod{n}\relwedge y\in\set{0,1,\ldots,n-1}}
  \end{equation}
\end{cor}
\begin{proof}
  For every~$r$ in~$\Z/n\Z$, \rproposition{pqrec.peri} yields that
  the set~$P_{n,\set{r}}$ is \pqrec,
  hence it follows from \rtheorem{pq-defi<->pq-reco}, that it
  is \pqdef by a formula~$\Psi_r(x)$.
  The relation of the statement is definable by the following formula.
  \begin{equation}
    \Phi(x,y) \defeq\quad\quad  \bigvee_{i=0}^{n-1}\Big( y\mathbin{=}i \relwedge \Psi_i(x)\Big)
  \end{equation}
  \rtheorem{pq-defi<->pq-reco} concludes the proof.
\end{proof}
%
%
%

\subsection{Equivalence modulo~$q$}
\lsection{modu-not-pqrec}

Unlike in the previous case, it is not obvious to extend the modulo-$q$ operator
to~$\Npq$.
However, as far as \pqrecy is concerned, the question is irrelevant:
any such generalisation is not \pqrec.
It is the \rcorollary{npqrec.modu} of the next statement.

\begin{prop}\lproposition{npqrec.modu}
Any subset~$S$ of~$\Npq$ such that~$S\cap \N = q\N$ is not \pqrec.
\end{prop}
\begin{proof} 
  For the sake of contradiction, we assume that there exists~$\Ac_0$, an
  automaton realising a set~$S\subseteq\Npq$ that satisfies~$S\cap
  \N = q \N$.
  \rtheorem{pq-defi<->pq-reco} yields that there exists, for every integer~$i$, $0\leq i < q$,
  an automaton~$\Ac_{i}=\langle \Ap, Q_i, \delta_i, i_i, F_i \rangle$ that realises~$S+i$.
  Note in particular that the~$\Ac_i$'s are correct if the input is the expansion of an integer:
  \begin{multline}\lequation{ai-corr-inte}
    \forall i,~0\mathbin{\leq} i \mathbin{<} q,\quad \forall u\in\Aps,~\val(u)\in\N,\\\Ac_i\text{ accepts } u \iff i=\val(u)\mod q~.
  \end{multline}
  %
  We assume without loss of generality that the~$\Ac_i$'s are
  complete and left-to-right.

  Let~$\Bc$ be the left-to-right automaton:
  \begin{equation*}
    \Bc = \langle\, \Ap,\,Q_\Bc,\,\delta_\Bc,\,i_\Bc,\, F_\Bc \,\rangle
  \end{equation*}
  where the state-set is~$Q_\Bc= Q_0\times \cdots \times Q_{q\mo}$, the initial state is~$i_\Bc=(i_0,\ldots,i_{q\mo})$,
  all states are final:~$F_\Bc=Q_\Bc$, and~$\delta_\Bc$ is defined as follows.
  Let~$\vect{s}=(s_0,s_1,\ldots,s_{q\mo})$ be a state in~$Q_\Bc$ and let~$a$ be a letter in~$\Ap$.
  \begin{multline}\lequation{important}
    \delta_\Bc(\vect{s},a)\quad\text{is defined if and only if}\quad s_i\in F_i~, \\
    \text{where }i\text{ is \underline{the} integer}\text{ such that}\quad i\xmd p+a \mathbin{\equiv} 0~[q]
  \end{multline}
In this case, $\delta_\Bc(\vect{s},a)=(
      \delta_i(s_0,a),  \delta_i(s_1,a), \ldots, \delta_i(s_{q-1},a))$.
%
  %

  %
  The important part of the definition of~$\Bc$ is \requation{important};
  it should be compared with~\requation{MEA-alt} (which gives the algorithm
  to compute the representation of integers).
  Intuitively, \requation{important} ensures that taking a transition in~$\Bc$
  preserves the property of \emph{being evaluated to an integer}; it will yield
  \rclaim{bc-corr} thanks to a proof by induction.
  The rest of the definition of~$\Bc$ is simply such that \rclaim{bc-comp} holds.

  \begin{claim}\lclaim{bc-comp}
    Let~$u$ be a word in~$\Aps$.
    If the run of~$u$ in~$\Bc$ exists and reaches the
    state~$(s_0,s_1,\ldots,s_{q\mo})$, then for every~$i$,~$0\leq i < q$, the
    run of~$u$ in~$\Ac_i$ exists and reaches~$s_i$.
    
  \end{claim}
  \begin{proofwithbar}
  From its definition, the automaton~$\Bc$ is the result of two transformation
  applied to the classical automaton product~$\Ac_0\times \Ac_1 \times \cdots \times \Ac_{q\mo}$:
  deleting some transitions and setting all states as final. Since the~$A_i$'s are all complete,
  Claim~\theclaim{} follows.
  \end{proofwithbar}

  \begin{claim}\lclaim{bc-corr}
    A word~$u$ has a run in~$\Bc$
    if and only if~$\val(u)$ is an integer.
  \end{claim}
  \begin{proofwithbar} By induction over the length of~$u$; the case~$u=\epsilon$ is trivial.

  Let~$u=v\xmd a$ be a non-empty word in~$\Aps$.  Induction hypothesis then rewrites as\hfill
  \noindent\begin{minipage}{\linewidth}
  \begin{subequations}
  \lequation{bc-corr-i}
  \begin{gather}
    \lequation{bc-corr-il}
    v\text{ has a run in }\Bc\\
    \Updownarrow \notag\\
    \lequation{bc-corr-ir}
    \val(v)\text{ belongs to }\N
  \end{gather}
  \end{subequations}
  \end{minipage}

  \medskip

  \noindent Note that~\emph{$u$ has a run in~$\Bc$} implies \requation*{bc-corr-il}, since~$v$ is a prefix of~$u$;
  and that~$\val(v\xmd a)\in \N$ implies~\requation*{bc-corr-ir}, since~$0^*\Lpq$ is a prefix-closed language.
  Hence, in both directions of the proof, both sides of the equivalence of \requation*{bc-corr-i} hold.
  Then, we let~$\vect{s}=(s_0,s_1,\cdots,s_{q\mo})$ denote the state reached by
  the run of~$v$ in~$\Bc$.
  Hence, for every integer~$i$, $0\leq i < q$, the run of~$v$ in~$\Ac_i$ 
  reaches~$s_i$ (\rclaim{bc-comp}).
  From~\requation*{ai-corr-inte}, since~$\val(v)$ is an integer,~$v$ is accepted by~$\Ac_{k}$, where~$k=\val(v)\mod{q}$,
  and rejected by each~$\Ac_i$ such that~$i\neq k$.
  %
  In other words,~$s_k\in F_k$ and for every~$i\neq k$,~$s_i\notin F_i$.
  Then, the following are equivalent.

  \smallskip

  \noindent \hspace*{5mm}$u$ has a run in~$\Bc$
  \begin{itemize}[label={\hspace*{5mm}${\iff}$},itemindent=5mm,itemsep=1mm]
    \item $\delta(\vect{s},a)=\vect{t}$ for some~$\vect{t}\in Q_\Bc$
    \item $s_i\in F_i$\quad and\quad$i\xmd p + a\equiv 0~[q]$ \quad(from \requation*{important})
    \item $k\xmd p + a \equiv 0~~[q]$\quad (since necessarily, $k\mathbin{=}i$)
    \item $\val(v)\xmd p+a \equiv 0~~[q]$\quad (from the definition of~$k$)
    \item $\frac{1}{q}(\val(v)\xmd p+a)\in\N$
    \item $\val(v\xmd a)=\val(u)\in\N$\quad (from \requation*{eval-recu})
  \end{itemize}

  \smallskip

  \noindent This concludes the proof of Claim \theclaim{}.
\end{proofwithbar}

  \medskip

  \noindent Since all states in~$\Bc$ are final, \rclaim{bc-corr} yields that the language accepted by~$\Bc$ is~$0^*\Lpq$,
  a contradiction.
\end{proof}
The inseparability result stated  by \rproposition{npqrec.modu} may be
generalised to every periodic set whose smallest period is a multiple of~$q$.

\begin{prop}\lproposition{insep}
  Let~$P$ be a periodic set of integers, the smallest period of which is a
  multiple of~$q$.
  Then, there is no \pqrec set~$S$ such that~$S\cap\N=P$.
\end{prop}
\begin{proof}
  Since bounded intervals are \pqrec (\rproposition{prec.inte}),
  we may assume that~$P$ is purely periodic.
  First, we assume moreover that the smallest period of~$P$ is exactly~$q$.
  In this case, there exists a remainder set~$R\subseteq \Z/q\Z$
  such that~$S=R+q\N$.
  Using that~$q$ is the \strong{smallest} period of~$P$, simple arithmetics shows that the following claim holds.

  \begin{claim}\lclaim{div}
    An integer~$i$ is divisible by~$q$ if and only if,
    for every~$r$ in~$R$, $(i+r)\mod{q}$ belongs to~$R$
  \end{claim}

  For the sake of contradiction, we assume that there exists a set~$S$
  as in the statement.
  We denote the formula that defines~$S$ by~$\Phi(x)$.
  Let~$\xi(x) \defeq\bigwedge_{r\in R}\Phi(x+r)$
  and let~$X$ be the set definable by~$\xi$.
  \rclaim{div} yields that ${\N\cap X =q\N}$,
  From \rtheorem{pq-defi<->pq-reco}, $X$ is \pqrec, a contradiction
  to \rproposition{npqrec.modu}.

  \medskip

  If the period of~$P$ is~$k\xmd q$,~$k>1$, a similar \emph{ab absurdo}
  reasoning yields a \pqrec set~$X$ such that~$\N\cap X =k\xmd q\xmd \N$.
  Then, using \rtheorem{pq-defi<->pq-reco}, it is easy to show that the set
  $Y=X \cup (X+q) \cup \cdots \cup (X+(k-1)\xmd q)$ is also \pqrec and
  satisfies~$Y\cap \N = q\xmd \N$, a contradiction to \rproposition{npqrec.modu}.
\end{proof}

\begin{cor}\lcorollary{npqrec.modu}
  Let~$k$ be a positive integer.
  Let~$f$ be a function~$\Npq\rightarrow\Npq$ such that, for every integer~$n$,~$f(n)=n\mod{k\xmd q}$.
  The function~$f$ is not \pqrec.
\end{cor}
In Sections~\rsection*{modu-pqrec}
and \rsection*{modu-not-pqrec}, we studied the \pqrecy of the modulo-$n$
operator when~$n$ is coprime with~$q$ (\rcorollary{pqrec.modu}) and when~$n$ is
a multiple of~$q$ (\rcorollary{npqrec.modu}).
These results do not cover all cases, and in general we conjecture the
following.

\begin{conj}
  Let~$n$ be an integer.
  Let~$f$ be a function~$\Npq\rightarrow\Npq$.
  We assume that, for every number~$\frac{x}{y}$ in~$\Npq$ such that~$y$ is
  coprime with~$n$, then $f(\frac{x}{y})= (x y^{-1})\mod{n}$,
  where~$y^{-1}$ is the inverse of~$y$ in~$\Z/n \Z$.
  Then, the function~$f$ is \pqrec if and only if~$n$ is coprime with~$q$.
\end{conj}

\section{Conclusion and future work}

In this work, we took a perspective which is classical for other numeration
systems (integer base, U-systems), but quite new in the studies on rational base
numeration systems.
Instead of exploring the intricacies of the language~$\Lpq$,
we indeed started to determine what may or may not be computed by automata.
It is very encouraging that the logic characterisation given by
\rtheorem{pq-defi<->pq-reco} is similar to the corresponding statements
in other settings \cite{BruyEtAl94,BruyHans97,Char18-ib}.
This opens for base~$\pq$ a lot of questions that have been answered for others
numerations systems.
For instance, it is natural to confront \pqrecy to sets definable in Presburger
arithmetic.
Indeed Cobham Theorem and its generalisations \cite{DuraRigo10} showed that a
set realised by automata in ``sufficiently different'' numerations systems is
necessarily definable in that logic.
In the case of base~$\pq$, the notion needs some adaptation, and
\rsection{nota} merely starts the research in that direction.
Another research direction would be to consider the fact that every real number
is represented in base~$\pq$
by an infinite word over~$\Ap$.
Then, one could study the subsets of~$\R^d$ that are realised by $d$-tape
B{\"u}chi automata and establish whether there is a logic characterisation of
them, much like what is done in~\cite{BoigEtAl98} for integer bases.
The considerable advantage would then be that the domain is always~$\R$,
independently of the base~$\pq$ considered.
Comparing \pqrecy to~$\frac{r}{s}$-recognisability could then lead to a statement in
the spirit of Cobham Theorem, similar to the result in~\cite{BoigEtAl10}.
%


\section*{Acknowledgements}
  Some of the results in this article are part of the Ph.D thesis of the author
  \cite{Mars16}, written under the supervision of Jacques Sakarovitch.
  The author warmly thanks Arnaud Carayol for several helpful discussions that
  made the logic characterisation in \rtheorem{pq-defi<->pq-reco} more elegant.
  The author is also grateful to the reviewers for their many valuable
  comments, which greatly improved the presentation of the results.

\nocite{BertRigo10-b,BertRigo18-b}
\bibliographystyle{alpha}
\bibliography{Alexandrie-abbrevs.bib,Alexandrie-AC.bib,Alexandrie-DF.bib,Alexandrie-GL.bib,Alexandrie-MR.bib,Alexandrie-SZ.bib,CANT.bib,bibliography.bib}
\end{document}